\documentclass[letterpaper,journal]{IEEEtran}
\usepackage{amsmath,amsfonts}
\usepackage{amsthm}
\usepackage{algorithm}
\usepackage{algpseudocode}
\usepackage{array}
\usepackage[caption=false,font=normalsize,labelfont=sf,textfont=sf]{subfig}
\usepackage{textcomp}
\usepackage{stfloats}
\usepackage{url}
\usepackage{verbatim}
\usepackage{graphicx}
\usepackage{cite}
\usepackage{wrapfig}
\usepackage{setspace}
\usepackage{import}
\usepackage[T1]{fontenc}
\usepackage[dvipsnames]{xcolor}
\usepackage{tikz}
\usetikzlibrary{arrows.meta, calc}

\usepackage{xifthen}
\usepackage{pdfpages}
\usepackage{transparent}

\usepackage[colorlinks=true, linkcolor=blue, urlcolor=blue, citecolor=blue, linkbordercolor=white]{hyperref}
\usepackage{enumitem}
\usepackage{blkarray}
\usepackage{amssymb}
\usepackage{tikz-cd}
\usepackage{aliascnt}
\usepackage{cleveref}
\usepackage{oubraces}

\theoremstyle{plain}
\newtheorem{theorem}{Theorem}[section]
\newaliascnt{lemma}{theorem}
\newtheorem{lemma}[lemma]{Lemma}
\aliascntresetthe{lemma}
\newaliascnt{corollary}{theorem}
\newtheorem{corollary}[corollary]{Corollary}
\aliascntresetthe{corollary}
\newaliascnt{proposition}{theorem}
\newtheorem{proposition}[proposition]{Proposition}
\aliascntresetthe{proposition}
\theoremstyle{definition}
\newaliascnt{definition}{theorem}
\newtheorem{definition}[definition]{Definition}
\aliascntresetthe{definition}
\newaliascnt{example}{theorem}
\newtheorem{example}[example]{Example}
\aliascntresetthe{example}
\newaliascnt{problem}{theorem}
\newtheorem{problem}[problem]{Problem}
\aliascntresetthe{problem}
\newaliascnt{conjecture}{theorem}

\aliascntresetthe{conjecture}
\theoremstyle{remark}
\newaliascnt{remark}{theorem}
\newtheorem{remark}[remark]{Remark}
\aliascntresetthe{remark}
\numberwithin{equation}{section}

\crefname{theorem}{theorem}{theorems}
\Crefname{theorem}{Theorem}{Theorems}
\crefname{lemma}{lemma}{lemmas}
\Crefname{lemma}{Lemma}{Lemmas}
\crefname{corollary}{corollary}{corollaries}
\Crefname{corollary}{Corollary}{Corollaries}
\crefname{proposition}{proposition}{propositions}
\Crefname{proposition}{Proposition}{Propositions}
\crefname{definition}{definition}{definitions}
\Crefname{definition}{Definition}{Definitions}
\crefname{example}{example}{examples}
\Crefname{example}{Example}{Examples}
\crefname{problem}{problem}{problems}
\Crefname{problem}{Problem}{Problems}
\crefname{conjecture}{conjecture}{conjectures}
\Crefname{conjecture}{Conjecture}{Conjectures}
\crefname{remark}{remark}{remarks}
\Crefname{remark}{Remark}{Remarks}

\newlength{\subfiguresize}
\newcommand{\R}{\mathbb{R}}
\newcommand{\bR}{\overline{\mathbb{R}}} 
\newcommand{\F}{\mathbb{F}_2} 

\newcommand{\e}{\epsilon}
\newcommand{\cl}[1]{\overline{#1}}

\newcommand{\bs}[1]{\boldsymbol{#1}}
\newcommand{\tree}{\mathfrak{T}}
\newcommand{\signal}{\mathcal{S}}
\newcommand{\PD}{\text{PD}}
\newcommand{\bht}{\text{BHT}}
\newcommand{\fmax}{f^{\uparrow}}
\newcommand{\gwf}{graph with faces}
\newcommand{\gwfs}{graphs with faces}
\newcommand{\G}{\mathcal{G}}
\newcommand{\Gs}{\bs{\G}}
\newcommand{\gs}{\bs{G}}
\newcommand{\lk}{L}
\newcommand{\emb}{\Phi}
\newcommand{\hole}{\mathcal{H}_\emb} 
\newcommand{\face}{\tau}
\newcommand{\lpf}{\mathcal{L}}
\newcommand{\deff}{\triangleq}
\newcommand{\ind}[1]{#1_{G[\emb]}}
\newcommand{\indf}{\ind{f}}
\newcommand{\lowerpd}{\lesssim}
\newcommand{\eqpd}{\approx}
\newcommand{\BD}{\text{BD}} 

\DeclareMathOperator{\Ima}{Im}
\DeclareMathOperator{\pers}{pers}
\DeclareMathOperator{\anc}{anc}

\def\withsupplement{}

\newif\ifincludesupplement
\includesupplementfalse
\ifdefined\withsupplement
  \includesupplementtrue
\fi

\begin{document}

\title{Topology-based Filtering of Graph Signals via Persistent Homology}

\author{Matias de Jong van Lier, Sebasti\'an El\'{\i}as Graiff Zurita and Shizuo Kaji
\thanks{This research was partially supported by JST Moonshot R\&D Grant Number JPMJMS2021 and
KAKENHI, Grant-in-Aid for Scientific Research (B) 25K00921 and (S) 25H00399.
}
\thanks{Matias de Jong van Lier is with the School of Mathematics, Kyushu University, Fukuoka 8190395, Japan (e-mail: de.jong.van.lier.matias.808@s.kyushu-u.ac.jp).
Sebasti\'an El\'{\i}as Graiff Zurita is with Graduate School of Science, Kyoto University, Kyoto 6068502, Japan (e-mail: s-graiff@kyudai.jp).
Shizuo Kaji is with Graduate School of Science, Kyoto University, Kyoto 6068502, Japan  (e-mail: kaji.shizuo.7r@kyoto-u.ac.jp).}
}



\maketitle

\begin{abstract}
We study topology-based filtering of vertex-defined signals on graphs and their two-dimensional analogues. Unlike graph-spectral filters, the proposed approach distinguishes features by topological persistence rather than by spatial wavelength or periodicity. We consider graphs with faces embedded in surfaces, a class that includes discrete models of images and meshes. We prove that, in general, exact simultaneous removal of low-persistence features in dimensions $0$ and $1$ is impossible. This motivates a relaxed formulation, for which we introduce the Low Persistence Filter (LPF). The LPF removes finite-persistence features below a prescribed threshold while controlling the resulting $\ell_\infty$ perturbation of the signal. We illustrate the method on one-dimensional signals, two-dimensional images, and signals on triangular meshes. A Python implementation is publicly available.
\end{abstract}


\setlist{itemsep=0.5em}

\section{Introduction}

Graph Signal Processing (GSP) studies how to analyse, denoise, and transform
signals supported on irregular domains such as sensor, transportation,
biological, and brain
networks~\cite{shuman2013,sandry2014,graph-signal-processing-overview}. Much of
the field is organised around graph-Laplacian notions of frequency---graph
Fourier analysis, spectral filters, wavelets, and filter banks---so that, as in
classical Fourier-based filtering, ``low'' and ``high'' features are defined by
frequency. Many of these methods are spectral or local in nature, whereas in
several applications the features of interest are intrinsically topological:
connected regions, \emph{basins} (valleys around local minima), and loops that
persist across scales.

This paper studies a complementary notion of low and high features. The
topological features we track are of two kinds: \emph{connected components} of
the sublevel sets of the signal (captured by $0$-dimensional homology) and
\emph{loops} in them ($1$-dimensional homology). We gauge the significance of
such a feature by its \emph{persistence}---how long it survives as the signal
level sweeps upward---and regard it as insignificant when it is not robust
under small perturbations of the signal, as quantified by persistent
homology~\cite{edel2002}, the standard multiscale tool of Topological Data
Analysis (TDA). Thus a spatially periodic ripple may be of low persistence,
while a localised but deep valley or loop may be of high persistence even
though it carries high graph-frequency content.

\begin{figure}
  \centering
  \includegraphics[width=0.98\linewidth]{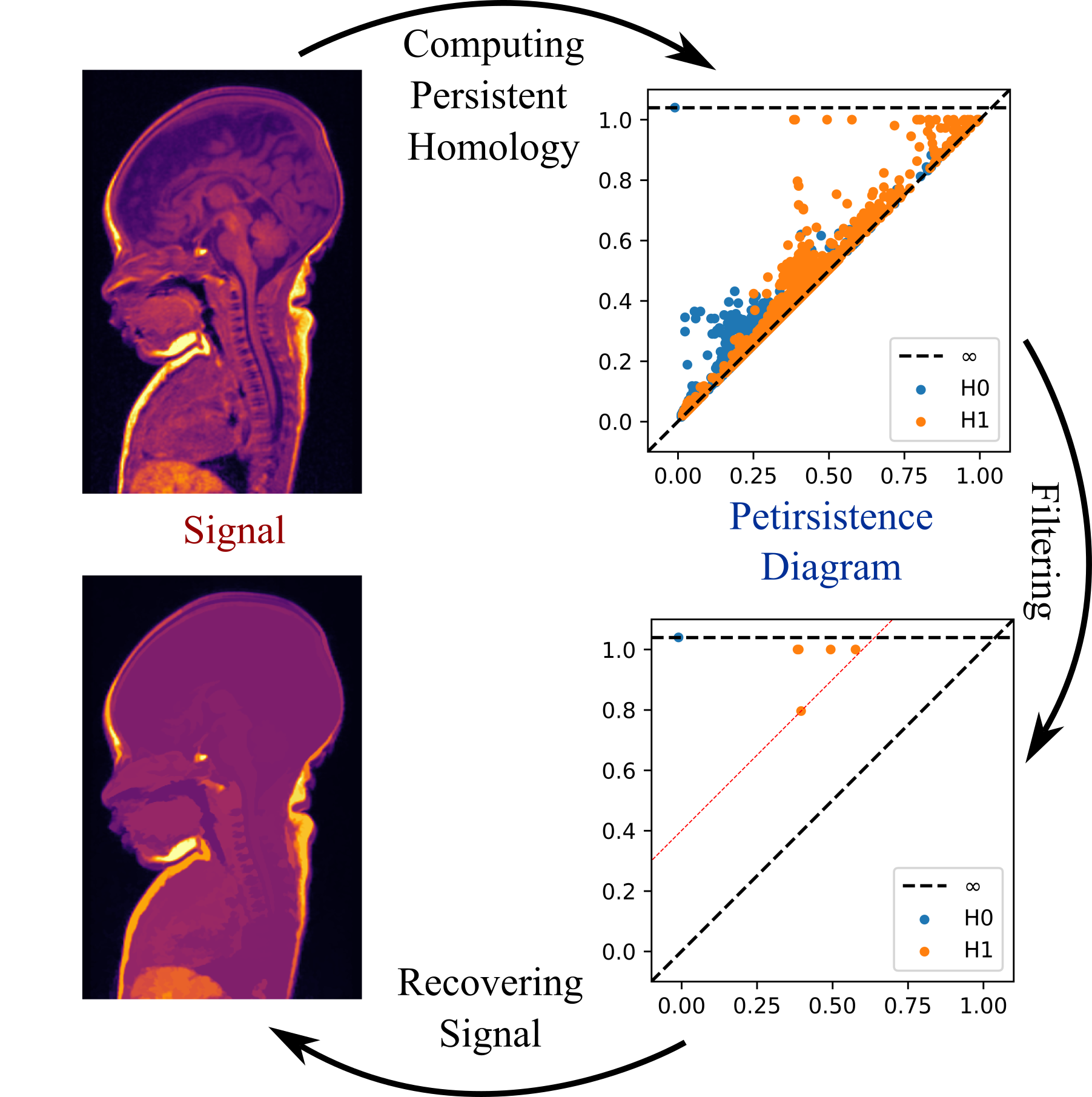}
  \caption{The LPF transforms the input image (top) into the
	processed image (bottom) by suppressing low-persistence topological
	features below the specified threshold (orange line) in the persistence
	diagram (right). The computation uses the publicly available Python
	implementation of the LPF. The original MRI image on
	the left is provided courtesy of Ian Bickle, Radiopaedia.org, rID: 52599.}
  \label{fig:mri-lpf-overview}
\end{figure}

We therefore ask the following question (\Cref{fig:mri-lpf-overview}): given a
vertex-defined signal on a fixed network, can one directly modify the signal
values so as to suppress topological features of small persistence without
changing the support itself? Here ``topological noise'' refers to features
deemed insignificant relative to an application-dependent threshold
$\epsilon$. Thus, unlike Laplacian-based filtering, which suppresses components
classified as spectrally high-frequency, our method suppresses features
classified as topologically short-lived.

For $0$-dimensional homology alone, this can often be done by lifting local
minima. When the same fixed-support, vertex-only modification must also control
$1$-dimensional loops, the two dimensions interact: filtering a
low-persistence basin can create or shift a loop, and conversely. 
We formulate this task as a topology-aware filtering problem for vertex-defined signals on generalized two-dimensional cell complexes embedded in a surface. We then show by example that exact simultaneous removal of all low-persistence features in dimensions $0$ and $1$ is impossible in general; this obstruction motivates a relaxed formulation and the Low Persistence Filter (LPF).
From a GSP perspective, the
LPF is a nonlinear filter whose selection rule is based on persistence rather
than frequency. 

To represent loops with finite persistence, we work with a higher-order
structure that we call a \emph{graph with faces}; without faces, every
$1$-homology class would be essential, of infinite persistence, so that no
multiscale analysis of loops would be possible. Unlike standard cell complexes,
our construction does not restrict its $2$-cells to disk-like regions, which
broadens the class of surface domains to which the filter applies.

\subsection{Related Work}\label{sec:related-work}

Classical graph-signal filtering is largely organised around the graph
Laplacian or related graph-shift operators. Early formulations of graph Fourier
analysis and linear graph filters on irregular domains appear
in~\cite{shuman2013,sandry2014,graph-signal-processing2,graph-signal-processing3,graph-signal-processing-overview}. Spectral
graph wavelets, filter banks, and more recent graph-filter overviews further
develop this Laplacian-based viewpoint for denoising, multiscale analysis, and
learning on fixed
graphs~\cite{graph-signal-processing4,graph-filters2,graph-filters3,graph-filters}. This
literature is the main signal-processing backdrop for our paper: as in these
works, we study filtering of signals on a fixed support. The distinction,
however, lies not only in the mathematical constructions, but also in the
underlying notion of noise: spectral methods typically suppress components
deemed noisy because they occupy high graph frequencies, whereas our method
suppresses features deemed noisy because they have low persistence. The two
criteria are complementary rather than interchangeable. Frequency-domain
methods, including FFT-based filters on regular grids, decompose signals
according to oscillation, while our filter changes vertex values according to
the persistence of sublevel-set features. The resulting filter is therefore
nonlinear and acts on persistence pairs rather than Laplacian frequencies.

Beyond vertex signals on graphs, higher-order GSP extends Laplacian ideas to
simplicial, cellular, and sheaf-based domains---beginning with Robinson's
sheaf-theoretic framework~\cite{robinson2014topological}, which generalises
attaching local signals to parts of a topological space---through Hodge
Laplacians and related operators~\cite{barbarossa2016, barbarossa2020,
  barbarossa2020b, schaub2020, schaub2021signal, Schaub2022, barbarossa2021,
  Roddenberry2022Cell, barbarossa2023, barbarossa2024}. These works show that
edges, faces, and higher-order relations support meaningful notions of
frequency, filtering, and signal decomposition. Our use of graphs with faces is
closer in spirit to this line, although our signal remains vertex-defined and
our goal is not a spectral decomposition on higher-order cells but direct
suppression of low-persistence topological features.

Persistent homology, the principal multiscale tool of TDA~\cite{edel2002}, has
been applied across biology, neuroscience, materials science, and signal
analysis~\cite{ghrist2008,TDA-carlsson,hiraoka2016}. Persistent Laplacians
connect persistence with Laplacian-based
operators~\cite{wang2020persistent,memoli2022persistent}, and persistent homology has been
widely used as a descriptor for structured data, including network and
brain-graph analysis~\cite{lee2011persistent, lee2012persistent}, time-series
signals~\cite{Perea2014, perea2015sw1pers, seversky2016time, Umeda2017,
  perea2019topological}, and recent graph-learning architectures based on
topological readouts or pooling~\cite{zhang2022gefl,
  chen2023topologicalpooling, ballester2025expressivity}. In these works,
topology is primarily used for representation, analysis, or 
characterisation. Our objective is different: we use persistence as the
criterion for directly constructing a filtered signal.

A line of work studies persistence-sensitive simplification of scalar functions
on manifolds and surfaces~\cite{edelsbrunner2006persistence,bauer2012optimal},
together with stability-based viewpoints that explain how low-persistence
features respond to perturbations~\cite{Edelsbrunner-stability-2005,
  Myers2022}; in particular, dedicated tools for persistent noise modelling and
thresholding for time series have been developed in the \textsc{Teaspoon}
library~\cite{khasawneh2025teaspoon}. These works help
motivate the persistence thresholding considered in this paper.

In the fixed-support, vertex-only setting, we prove an impossibility result for
exact simultaneous removal of low-persistence features in dimensions $0$ and
$1$, formulate a relaxed problem, and give a construction with an explicit
$\ell_\infty$ bound.

\subsection{Contributions and organisation}

The contributions and organisation are as follows:
\begin{itemize}
\item \textbf{Higher-order domain (\Cref{sec:gwf}):} We introduce \emph{graphs
with faces} as a higher-order domain for vertex-defined signals, embedded in an
  arbitrary connected closed surface. We prove that an induced graph computes
  the complete $0$-dimensional diagram, all finite-persistence
  $1$-dimensional intervals, and the number of essential handle intervals
  (\Cref{thm:induced-graph-PH}).
\item \textbf{Problem formulation and obstruction
  (\Cref{sec:original-problem}):} We treat topology-based filtering as a
  fixed-support filtering problem where low-persistence features are considered
  noise. We show that, in general, one cannot simultaneously eliminate all
  subthreshold $\PD_0$ and $\PD_1$ intervals while preserving every interval
  above the threshold.
\item \textbf{BHT representation and LPF (\Cref{sec:bht}):} We introduce the
  \emph{Basin Hierarchy Tree}, which records the representative minimum,
  parent, and linking vertex of each basin. These data are used to define the
  LPF, whose alternating construction satisfies the approximation bound in
  \Cref{thm:lpf-all}.
\item \textbf{Examples (\Cref{sec:experiments}):} We compare persistence-based
  and frequency-based filtering on synthetic signals and apply the LPF to a
  natural image and a triangular mesh signal.
\end{itemize}

All proofs are given in the Supplemental Material.

\section{Preliminaries}

\subsection{Graph with faces}\label{sec:gwf}

Several notions of \emph{faces} on graphs appear in the literature. Familiar
examples are the $2$-cells of simplicial and cubical
complexes~\cite{algtop-Hatcher}, which are respectively triangles and squares.
A more flexible model is the \emph{2-cell
embedding}: a graph embedding $\gamma: G \hookrightarrow \Sigma$ into a closed
surface $\Sigma$ such that each connected component of $\Sigma \setminus
\gamma(G)$ is homeomorphic to an open disk. These components are the
\emph{faces}. Although more general than simplicial or cubical complexes, this
notion still restricts all faces to be disk-like. The Planar Hollow Cell
Complex (PHCC)~\cite[\S II, Def. 3]{barbarossa2024} generalises further: faces
are open disks with finitely many disjoint closed disks removed, and outer
boundaries homeomorphic to a circle $S^1$.

For the filtering problem studied here, we work with a slightly broader notion
of faces on surface-embedded graphs. We retain a graph embedding $\gamma: G
\hookrightarrow \Sigma$ into a closed surface, but treat \emph{any} connected
component of the complement $\Sigma \setminus \gamma(G)$ as a face, regardless
of topology. This definition includes the simplicial, cubical, and cell-complex
settings used in our applications and is expressed algebraically through the
boundary operator.

\begin{definition}\label{def:graph-with-faces}
A \textbf{face} of a graph $G = (V, E)$ is a recorded boundary $1$-cycle,
namely a finite list of edge occurrences \( \face = e_1 \dots e_N, \quad e_i
\in E, \) satisfying
\[
    \sum_{i=1}^N \partial_1(e_i) = 0,
\]
where $\partial_1(uv) = u + v$ and arithmetic is in $\F$. That is,
$\sum_{i=1}^N e_i$ is a $1$-cycle.  The recorded list, including repeated edge
occurrences, is part of the face data; incidence notation such as \(e\in\face\)
refers to its support, meaning that \(e\) occurs at least once in the list.

A \textbf{graph with faces} is a triple $\G = (V, E, F)$, where $F$ is a set of
such faces.  We call an element $\sigma\in V\cup E\cup F$ a \textbf{cell}.
\end{definition}

When $F = \emptyset$ (the empty set), a \gwf\ reduces to a standard graph
$G=(V,E)$.  A signal over either object is written $\gs=(G,f)$ or $\Gs=(\G,f)$,
where $f:V\to\R$ is vertex-defined.  For $\sigma\in V\cup E\cup F$, we write
$V(\sigma)$ for its incident vertex set; in particular, $V(v)=\{v\}$ for
vertices, while $V(e)$ is the set of endpoints of an edge $e$. We then set
\[
\fmax(\sigma) \deff \max f\bigl(V(\sigma)\bigr),
\]
so $\fmax(v)=f(v)$. We also extend $f$ to the auxiliary symbol $\emptyset$ by
the convention $f(\emptyset)=+\infty$, which is used to encode an
essential branch in the BHT.

A \textbf{$\Gs$-ordering} on $V \cup E \cup F$ is a total order $\prec$ such
that:
\begin{align}
  \sigma_1 \prec \sigma_2 &\Rightarrow \fmax(\sigma_1) \leq \fmax(\sigma_2), \label{eq:g-ord-vert} \\
  v \prec e               &\text{ for all } v \in V(e), \ e\in E\label{eq:g-ord-edge} \\
  e \prec \face           &\text{ for all } \face\in F \text{ and } e\in\face. \label{eq:g-ord-face2}
\end{align}
Thus a $\Gs$-ordering is nondecreasing in the cell values and, within each
level, places every vertex before its incident edges and every edge before its
incident faces.

To connect our abstract definition to geometric models, we consider graphs
embedded in a surface.

\begin{definition}\label{def:embedded-gwf}
Let $\Sigma$ be a connected closed surface. A \gwf\ $\G = (V, E, F)$ is \textbf{embedded} in $\Sigma$ if $(V, E)$ admits an embedding $\gamma: (V, E) \hookrightarrow \Sigma$ and there exists an injective map $\phi$ from $F$ to the set of connected components of $\Sigma \setminus \Ima(\gamma)$ such that
\[
    \partial \phi(\face) = \bigcup_{e \in \face} \gamma(e).
\]
The pair $\emb = (\gamma, \phi)$ is a \textbf{surface embedding} of $\G$, and $\Sigma$ is the \textbf{ambient surface}. The \textbf{holes} of $\G$ are the components of $\Sigma \setminus \Ima(\gamma)$ not in $\Ima(\phi)$, denoted $\hole$. The \textbf{embedded graph} is
\[
    \G_\emb := \gamma((V,E)) \cup \bigcup_{\face \in F} \overline{\phi(\face)} = \Sigma \setminus \hole \subseteq \Sigma.
\]
When $\Sigma = S^2$, it is the usual planar embedding.\footnote{Since $S^2$ minus a point is homeomorphic to $\mathbb{R}^2$, embeddings into $S^2$ or $\mathbb{R}^2$ are equivalent.} 
\end{definition}

Thus $F$ specifies the complement components treated as faces, and the
remaining complement components form $\hole$.

\begin{figure}[ht]
    \centering
    \includegraphics[width=1\linewidth]{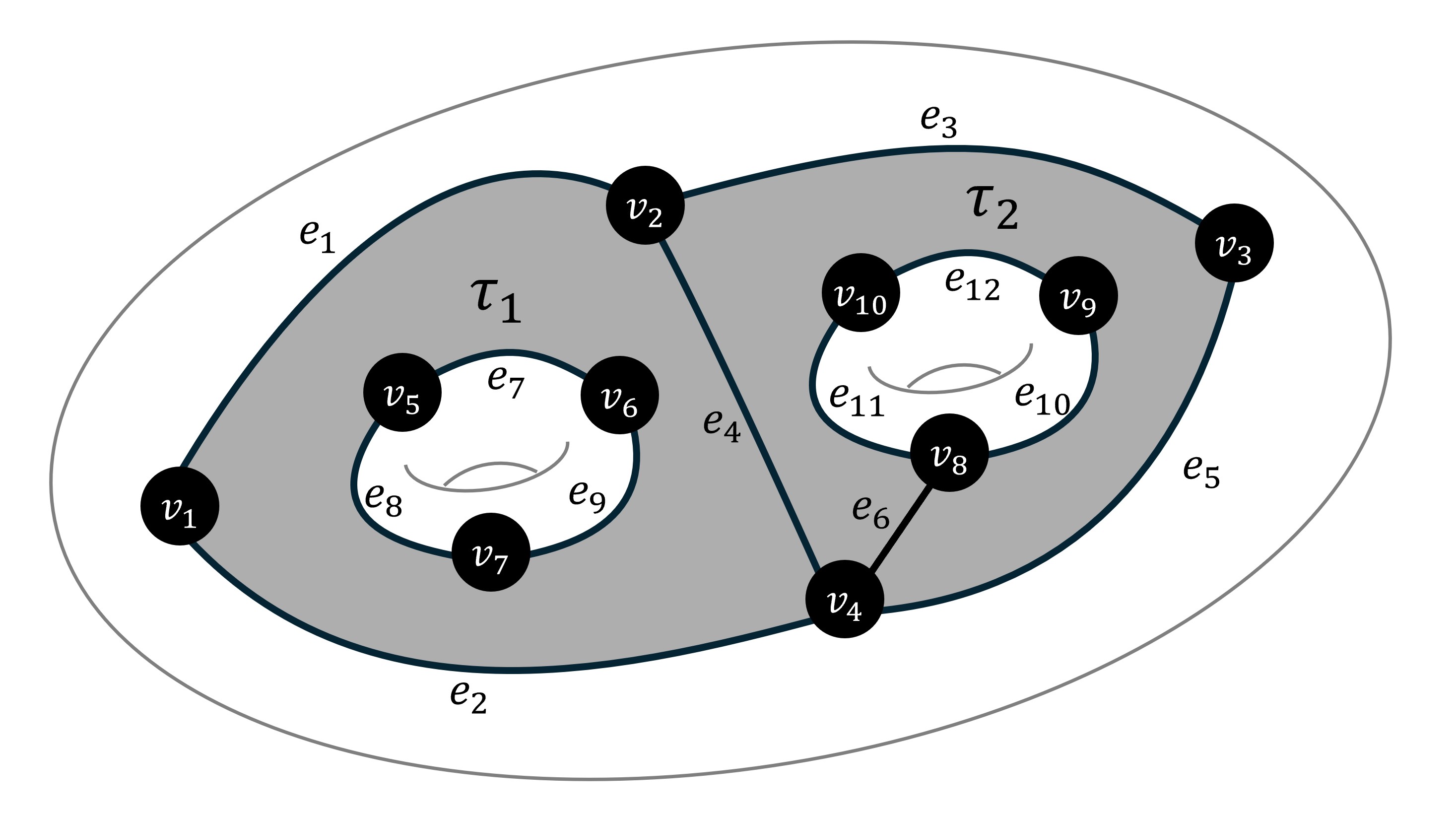}
    \caption{Example of a \gwf\ embedded into a genus-two surface. This is neither a 2-cell embedding nor a planar hollow cell complex. }
    \label{fig:non-phcc-gwf}
\end{figure}

\begin{example} \label{ex:gwf-not-phcc}
\Cref{fig:non-phcc-gwf} shows an embedded \gwf\ with \( H_0(\G)=\F \), \( H_1(\G) = \F^2 \). The face \( \face_1 = e_1e_2e_4e_7e_8e_9 \) is not homeomorphic to a disk, so this is not a 2-cell embedding. The face \( \face_2 = e_3e_4e_6e_{11}e_{12}e_{10}e_6e_5 \) has a boundary that is not homeomorphic to \( S^1 \), and thus fails the PHCC conditions~\cite[\S II, Def. 3]{barbarossa2024}.
\end{example}

\begin{example}
A greyscale image defines a signal over a planar \gwf\ based on a square grid graph. Vertices correspond to pixels; unit squares form the faces; greyscale values define the signal. 
\end{example}

\subsection{Homology}\label{sec:homology}
Homology assigns to a space $\mathcal{A}$ a sequence of vector spaces
$H_0(\mathcal{A}),H_1(\mathcal{A}),\dots$; only the first two are used in this
paper. Informally, $\dim H_0(\mathcal{A})$ is the number of \emph{connected
components}, and $\dim H_1(\mathcal{A})$ is the number of \emph{independent
loops} (one-dimensional holes that are not filled in) of $\mathcal{A}$. These
numbers are called the \emph{Betti numbers}, $b_i(\mathcal{A})=\dim
H_i(\mathcal{A})$. We use coefficients in the field with two elements $\F =
\{0,1\}$, so that homology is computed by linear algebra modulo $2$ and no
orientations are needed; readers wanting a self-contained, example-driven
account may consult the Supplemental Material.

The homology of a \gwf\ $\G$ embedded in $\Sigma$ is defined as the singular
homology of its embedded space $\G_\emb = \Sigma \setminus \hole$. Writing $c$
for the number of connected components of $\G_\emb$ and $b_1(\Sigma) =
\dim_{\F} H_1(\Sigma)$ for the first Betti number of the ambient surface, we
have $H_0(\G) = \F^c$ and, for a connected $\G_\emb$ with at least one hole,
\begin{align}\label{eq:H1}
  \dim_{\F} H_1(\G) = \bigl(|\hole|-1\bigr) + r, \\
  r \deff \dim_{\F}\operatorname{im}\bigl(H_1(\G_\emb)\to H_1(\Sigma)\bigr) \le b_1(\Sigma),
\end{align}
and $H_1(\G) = \F^{\,b_1(\Sigma)}$ when $\hole = \emptyset$ (so $\G_\emb =
\Sigma$). The summand $|\hole|-1$ counts the cycles that \emph{separate}
$\Sigma$, i.e.\ those bounding a hole, and $r$ counts the non-separating
\emph{handle} cycles; for disk-like holes $r = b_1(\Sigma)$. The planar case
$\Sigma = S^2$ has $b_1(S^2) = 0$ and $r = 0$.

\subsection{Persistent homology of a \gwf}\label{sec:ph}

Persistent homology extends homology from a single space to a signal-induced
family of spaces.  It records how homology evolves while a level $t$ sweeps
upward through the sublevel sets $\{\fmax \le t\}$: connected components and
loops are \emph{born} and later \emph{die}, and each feature is summarised by
the half-open interval $[b,d)$ between its birth level $b$ and death level
$d$. The length $d-b$ is its \emph{persistence}, and the multiset of all such
intervals in dimension $i$ is the \emph{persistence diagram} $\PD_i$;
features that never die (such as the component that remains once everything
has merged) are \emph{essential} and recorded with $d=+\infty$. We make this
precise below; the Supplemental Material gives an expanded, example-driven
version of the same construction.

Let $\Gs=((V,E,F),f)$ be a signal over a \gwf\ with a $\Gs$-ordering
$\prec$. From now on, we focus on the surface-embedded case, with embedding
data $\emb=(\gamma,\phi)$ into a closed surface $\Sigma$ for $\G$.

Define $\G_{\prec \sigma}$ and $\G_{\preceq \sigma}$ as the \gwfs\ having all
vertices, edges and faces coming before $\sigma$ in the ordering, including
$\sigma$ for the latter. Define $\G_t$ as the \gwf\ having all $\sigma\in V\cup
E\cup F$ such that $\fmax(\sigma)\leq t$.  With $\bR = \R \cup \{\pm\infty\}$
denoting the extended real line, $\{\G_t\}_{t\in\bR}$ is the
\textbf{sublevel filtration} of $\Gs$. Its persistent homology is represented
by the persistence diagrams \(\PD_i(\G,f)\), whose elements are intervals in
$\bR$.

In the sublevel filtration, if a homology class in dimension $i$ is born when
$\sigma_b$ appears and dies when $\sigma_d$ is included, it gives
rise to the interval \([f^\uparrow(\sigma_b), f^\uparrow(\sigma_d))\)
in the $i$-dimensional \emph{persistence diagram} \(\PD_i(\G, f)\). The pair
$(\sigma_b, \sigma_d)$ is called the \emph{birth-death pair} associated with
this interval. Birth--death pairs depend on the ordering when equal values
occur; we therefore denote by $\BD(\G, \prec)$ the set of all finite-death
birth--death pairs across all homological dimensions computed with respect to
the fixed $\Gs$-ordering \(\prec\). Unlike displayed diagrams, $\BD(\G,\prec)$
retains finite-death zero-persistence pairs; these formal pairs are needed for
the bijections in \Cref{def:lower-pd}. Essential intervals are
included in the persistence diagrams but not in $\BD(\G,\prec)$ as they have no
finite death cell.  The reduced $0$-dimensional persistence diagram
$\widetilde{\PD}_0$ is obtained by removing an essential interval with the
smallest birth.

The persistence module defined by the sublevel filtration is independent of
the $\Gs$-ordering. Consequently, diagrams computed from different admissible
orderings agree after zero-persistence intervals $[a,a)$ are omitted. We
denote equality after omitting these intervals by $\eqpd$.

We next describe an induced-graph construction that computes both \(\PD_0\) and \(\PD_1\) of an embedded graph with faces by ordinary \(0\)-dimensional persistence on a graph.
The method for computing persistent $1$-homology is based on Alexander duality in the ambient surface \cite[Theorem 3.43]{algtop-Hatcher}---a classical duality relating the loops of a region to the connected components of its complement; for the planar case $\Sigma = S^2$ this is the classical statement, and on a surface of positive genus it acquires a correction (\Cref{thm:induced-graph-PH}).

\begin{definition}\label{def:graph-ind-emb}
    Let $\G = (V,E,F)$ be a \gwf\ embedded in a surface $\Sigma$ with surface embedding $\emb$. Define the graph $G[\emb] =
    (V',E')$, where $V' = V\cup F\cup \hole$ and
    $E' = E\cup E_F \cup E_{\hole}$. Here, $E_F$ consists of all edges connecting each face to its incident vertices:
    \begin{align}
    E_F \deff \Big\{ (\face, v)\ \Big|\ \face \in F \text{ and } v \in V(\face) \Big\},
    \end{align}
    and $E_{\hole} \deff \{ (h, v) \mid h\in\hole,\ \gamma(v)\in\partial h \}$ is defined analogously, using the vertices on the topological boundary of each hole.
    Furthermore, if $f$ is a signal over $\G$, we define the
    \textbf{signal induced on $G[\emb]$ by $f$} as 
    \begin{align}\label{eq:induced-signal}
        \indf(v) \deff \begin{cases}
            \fmax(v), &\text{ if $v\in V\cup F$},\\
            +\infty, &\text{ if $v\in \hole$}.
        \end{cases}
    \end{align}
\end{definition}

\Cref{fig:gwf-induced-graph} shows this construction for a planar example. The
auxiliary value $\infty$ assigned to hole vertices places them after all
finite-valued cells in the extended order.

\begin{figure}[ht]
    \centering
    \includegraphics[width=1\linewidth]{fig/gwf-induced-graph.pdf}
    \caption{A planar \gwf\ (left) and the induced graph (right).}
    \label{fig:gwf-induced-graph}
\end{figure}

\begin{theorem}\label{thm:induced-graph-PH}
    Let $\Gs = (\G, f)$ be a signal over a \gwf\ embedded in a connected closed surface $\Sigma$ with surface embedding $\emb$. Then
    \begin{align}
        \PD_0(\G, f) \eqpd \PD_0(G[\emb], \indf),
    \end{align}
    and, applying the parameter reversal $[a,b) \leftrightarrow [-b,-a)$ to the first summand below,
    \begin{align}\label{eq:pd1}
        \PD_1(\G, f) \;\simeq\; \widetilde{\PD}_0(G[\emb], -\indf)\ \sqcup\ \mathcal{E}_\Sigma,
    \end{align}
    where $\mathcal{E}_\Sigma$ consists of exactly
    $r$ essential ($+\infty$-persistence) intervals, where $r$ is defined in
    \Cref{eq:H1}. In particular, the reversal is a bijection on the
    finite-persistence intervals of the two diagrams. The dual construction
    determines the number, but not the birth levels, of the intervals in
    $\mathcal{E}_\Sigma$; a finite-threshold LPF never selects these intervals,
    although on a positive-genus surface their birth levels need not remain
    fixed.
\end{theorem}

\section{The Low Persistence Filtering Problem}\label{sec:original-problem}

By \emph{topological noise}, we mean finite persistence intervals below the user-specified threshold \(\epsilon\). The most stringent filtering requirement is to remove all topological noise below $\epsilon$ while keeping every interval above the threshold intact. We formalise this requirement first.

For any multiset $M$ consisting of intervals in $\bR$, we define the following submultisets
\begin{align}
    &M_{\geq \epsilon} = \{[a,b)\in M\ |\ b-a\geq\e \} \label{eq:PDgeq},\\
    &M_{< \epsilon} = \{[a,b)\in M\ |\ b-a<\e \} \label{eq:PDlower}.
    \end{align}
The submultiset $M_{>\epsilon}$ is defined analogously. We say two multisets $M$ and $M'$ are equivalent, denoted $M\eqpd M'$, when they agree after trivial intervals are discarded; that is, $M_{>0} = M'_{>0}$. We write $M^{\mathrm{fin}}$ for the submultiset of intervals with finite death.

\begin{problem}\label{prb:graph}
Let $\signal(\G)$ be the set of all signals $f:V\to\R$ over $\G=(V,E,F)$. 
Given a signal over a \gwf\ $\Gs=(\G,f)$ and a threshold $\e>0$, find a signal $g\in\signal(\G)$ over the same \gwf\ with $\|f-g\|_\infty < \e$ such that
    \begin{align}
        \PD_n(\G,g) \eqpd \PD_n(\G,f)_{\geq \e} \quad (n=0,1). \label{eq:LPF0}
    \end{align}
\end{problem}

The following diagram summarises the desired lifting from persistence-diagram thresholding to signal filtering:
\begin{center}
\begin{tikzpicture}
    \coordinate (A) at (0, 0);
    \coordinate (B) at (6, 0);
    \coordinate (C) at (6, 2);
    \coordinate (D) at (0, 2);

    \def\offseth{0.2}
    \def\offsetv{0.2}

    \draw[->] ($(A)!\offseth!(B)$) -- node[below=6pt] {Filtering the persistence diagram} ($(B)!\offseth!(A)$);
    \draw[->] ($(B)!\offsetv!(C)$) -- node[left] {Lifting $\PD_n$} ($(C)!\offsetv!(B)$);
    \draw[dashed,->] ($(D)!\offseth!(C)$) -- node[above] {Filtering the function} ($(C)!\offseth!(D)$);
    \draw[->] ($(D)!\offsetv!(A)$) -- node[left] {$\PD_n$} ($(A)!\offsetv!(D)$);

    \node at (A) {$\PD_n(\G,f)$};
    \node at (B) {$\PD_n(\G,f)_{\geq\e}$};
    \node at (C) {$(\G,g)$};
    \node at (D) {$(\G,f)$};
\end{tikzpicture}
\end{center}

By the stability theorem for persistence diagrams, a perturbation bounded by
\(\delta\) in the supremum norm changes the diagram by at most \(\delta\) in
bottleneck distance~\cite{Edelsbrunner-stability-2005}. \Cref{prb:graph} asks
for the converse type of operation: remove the intervals below \(\e\) by
changing only vertex values, while keeping the support fixed and preserving
the intervals of persistence at least \(\e\).

\subsection{Counterexample}
The $n=0$ instance of \Cref{prb:graph} is solvable on every embedded \gwf, and
the $n=1$ instance is solvable whenever the embedded space carries no essential
ambient-surface handle class (in particular, in the planar case). On an
arbitrary closed surface, the construction in \Cref{sec:lpf} removes the
subthreshold finite-persistence $1$-dimensional intervals but need not preserve
the birth levels of essential handle intervals. Even in the planar case, a
single signal $g$ need not satisfy the requirements in both dimensions.
The following example shows that this is impossible in general.

\begin{figure}[ht]
    \centering
    \includegraphics[width=\linewidth]{fig/counterexample-both.pdf}
    \caption{Example of a signal over a \gwf\ with no solution to \Cref{prb:graph}.}
    \label{fig:counterexample}
\end{figure}

\begin{example}\label{ex:counter-example}
    Let $\Gs = (\G, f)$ be the signal over a planar \gwf\ shown in \Cref{fig:counterexample}. Its persistence diagrams are
    \begin{align*}
        &\PD_0(\G,f) = \{[-3,\infty), [1,2)\},\\
        &\PD_1(\G,f) = \{[1,6), [2,9)\}.
    \end{align*}
    If we choose $\e = 2$, the possible $g:V\to\R$ with
    \begin{align}
        \|f-g\|_\infty < \e = 2 \label{eq:example-bound}
    \end{align}
    are signals as in the right side of \Cref{fig:counterexample}, with
    \begin{alignat*}{4}
        &w\in (-5,-1),\quad &&y\in(4,8),\quad &z\in (7,11)\\
        &x_1,x_2 \in (-1,3),\quad &&x_3 \in (0,4). &{}
    \end{alignat*}
    Suppose such values satisfy \Cref{eq:LPF0}.
    Preserving $[-3,\infty)\in \PD_0$ forces $w=-3$.
    Eliminating $[1,2)\in \PD_0$ forces $x_1\ge x_2\ge x_3$.
    Preserving $[1,6)\in \PD_1$ forces $y=6$ and $x_1=1$.
    Preserving $[2,9)\in \PD_1$ forces $z=9$ and $x_2=2$, contradicting $x_1\ge x_2$.
\end{example}

\subsection{Revised problem}\label{sec:revised-problem}
A first approach is to weaken condition~\eqref{eq:LPF0} to
    \begin{align}\label{eq:prb2}
        \PD_n(\G,g)_{<\e} \eqpd \emptyset, \quad n=0,1
    \end{align}
in order to permit the filtered signal to contain intervals of persistence at
least \(\e\) that are absent from the original signal. We exclude the creation
of such intervals by imposing the following monotonicity condition on
birth--death pairs.

\begin{definition}
  \label{def:lower-pd}
  Let $f_1$ and $f_2$ be signals on a \gwf\ $\G$.  We write $f_2 \lowerpd' f_1$
  when there are a $(\G, f_1)$-ordering $\prec_1$, a $(\G, f_2)$-ordering $\prec_2$,
  and a bijection
  \[
  \Psi\colon \BD(\G, \prec_1) \longrightarrow \BD(\G, \prec_2)
  \]
  that preserves homological dimension and fixes the \emph{representative
  cell} of each birth--death pair: if
  $(\sigma_b',\sigma_d')=\Psi(\sigma_b,\sigma_d)$, then
  $\sigma_b'=\sigma_b$ in dimension $0$ and $\sigma_d'=\sigma_d$ in dimension
  $1$. The bijection must also not increase persistence; that is,
  \[
  \big|f_2^\uparrow(\sigma_b')- f_2^\uparrow(\sigma_d')\big|
  \leq
  \big|f_1^\uparrow(\sigma_b)- f_1^\uparrow(\sigma_d)\big|.
  \]
  The transitive closure is denoted by $f_2 \lowerpd f_1$.
\end{definition}

Concretely, the representative cell has a basin interpretation. In dimension $0$ it is the local minimum of a basin---the birth cell of the corresponding interval. In dimension $1$ it is the cell that fills the loop---the death cell---which the dual construction of \Cref{thm:induced-graph-PH} identifies with the birth cell of a $0$-dimensional feature of $-\indf$. 

Thus, imposing $g\lowerpd f$ allows the finite-persistence topology only to be
simplified---representative features may be shortened or removed, but not
replaced by genuinely new high-persistence features.

The relaxed problem is therefore as follows.
\begin{problem}\label{prob:reformulated}
    Given a signal over a \gwf\ $\Gs=(\G,f)$ and a threshold $\e>0$, find $g\in\signal(\G)$ such that $\|f-g\|_\infty < \e$, $\PD_n(\G,g)_{<\e} \eqpd \emptyset$ for $n=0,1$, and $g\lowerpd f$.
\end{problem}

\begin{example}\label{ex:optimal}
    Let $(\G,f)$ be as in \Cref{fig:optimal-case-eps2} and consider a solution $g:V\to\R$ to \Cref{prob:reformulated} with $\e > 1$.
    Since ${\PD_0(\G,f) \eqpd \{[-5,\infty), [1,2)\}}$ and ${\PD_1(\G,f)=\{[2,3)\}}$,
    we must have ${\PD_0(\G,g) \eqpd \{[g(v_0),\infty)\}}$ and ${\PD_1(\G,g) \eqpd \emptyset}$.
    From the condition for $\PD_0$, we must have
    \begin{gather}
        g(v_0) = \min g(V), \quad g(v_2)\leq g(v_1).\label{eq:min1} \\
        \intertext{Similarly, to eliminate $[2,3)\in\PD_1(\G,f)$, we must have}
        g(v_3)\leq g(v_2).\label{eq:min3}
    \end{gather}
    Any optimal solution $g$ minimising ${\|f-g\|_\infty}$ subject to \Cref{eq:min1,eq:min3} satisfies $g(v_0)\in[-6,-4]$ and $g(v_1)=g(v_2)=g(v_3)=2$.
    Therefore, $\|f-g\|_\infty = 1$.
    Since this is valid for any $\e>1$, the upper bound $\|f-g\|_\infty <\e$ in \Cref{prob:reformulated} is sharp.

    \begin{figure}[ht]
        \centering
        \large
        \includegraphics[width=0.65\linewidth]{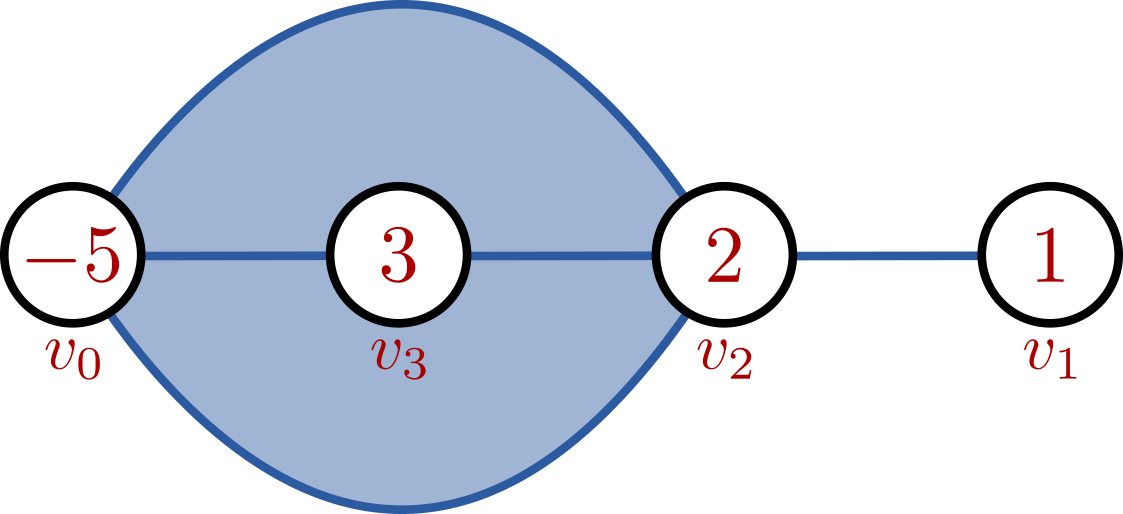}
        \caption{Example of a signal over a \gwf\ achieving an optimal approximation error.}
        \label{fig:optimal-case-eps2}
    \end{figure}
\end{example}

\Cref{thm:lpf-all} gives an explicit solution to
\Cref{prob:reformulated}. We first define the filters for each homological
dimension.

\section{Proposed Method}

The Low Persistence Filter (LPF) uses the induced graph of
\Cref{def:graph-ind-emb} as follows:
\begin{enumerate}
    \item Given a signal over an embedded \gwf, $(\G=(V,E,F), f:V\to\R)$, and surface embedding data $\emb=(\gamma,\phi)$, construct the induced graph along with the induced signal, $(G[\emb], \indf:V'\to\bR)$.
    \item Apply the graph $0$-LPF to $\indf$, eliminating its finite
    $0$-dimensional intervals of persistence below $\e$.
    \item Apply the same graph filter to the dual signal, again eliminating $0$-dimensional homology below the threshold $\e$. By \Cref{thm:induced-graph-PH}, this step corresponds to eliminating $1$-dimensional homology in the original signal.
    \item Restrict the resulting induced signal back to the original vertex set \(V\); the face values are then re-induced from these updated vertex values.
\end{enumerate}
Steps 2 and 3 are alternated because a vertex-value change in one homological
dimension can alter the persistence diagram in the other. The finite
termination and the resulting bounds are stated in \Cref{thm:lpf-all}.

\subsection{The Basin Hierarchy Tree}\label{sec:bht}


Before presenting our filtering method, we introduce the \emph{Basin Hierarchy Tree} (BHT), a representation that records how basins of a vertex-defined signal merge. Its construction follows the same union-find logic that underlies standard algorithms for $0$-dimensional persistent homology~\cite[Ch.~V.4]{TDA-Edelsbrunner}. The role of the BHT here is more specific: it retains the representative minimum, parent basin, and linking vertex needed to modify the original signal values while preserving control of the persistence pairs.

Throughout this subsection, we consider signals over connected ordinary graphs; the construction is later applied to the induced graph of a \gwf\ through \Cref{thm:induced-graph-PH}.\footnote{This entails no loss of generality: even when the underlying graph of a \gwf\ is disconnected, the induced graph remains connected; see the Supplemental Material.}
We employ standard graph-theoretic terminology. A \textbf{rooted tree} $T = (V, E_T)$ is a tree with a designated root vertex $r$. For any non-root vertex $v$, its \textbf{parent} $p(v)$ is the unique neighbour on the path to $r$. For a graph $G=(V,E)$, we denote by $C(v,G)\subset V$ the vertex set of the connected component of $G$ containing $v$.

\begin{proposition}\label{thm:BHT}
Let \(\gs=((V,E),f)\) be a signal over a connected graph, and let $\prec$ be a $\gs$-ordering. Write \(r=\min V\), where the minimum is taken with respect to $\prec$. For each \(v\in V\setminus\{r\}\), there is a unique edge \(m(v)\in E\), called the \textbf{merge edge of \(v\)}, such that
    \begin{align}
        \min C(v,G_{\prec m(v)}) &= v, \label{eq:merge-edge-before}\\
        \min C(v,G_{\preceq m(v)}) &\prec v. \label{eq:merge-edge-after}
    \end{align}
Set \(p(r)=r\) and \(p(v)=\min C(v,G_{\preceq m(v)})\) for \(v\neq r\). Then
\[
    T=(V,\{\{v,p(v)\}\mid v\in V\setminus\{r\}\})
\]
is a rooted tree with root \(r\). Since the merge edge \(m(v)\), and hence the parent \(p(v)\), is unique for every \(v\neq r\), this rooted tree is uniquely determined.
It is computed by \Cref{alg:bht}.
\end{proposition}
\begin{proof}
For a fixed \(v\neq r\), the edge \(m(v)\) is the first edge in the
$\gs$-ordering at which the minimum of the component containing \(v\) becomes
strictly smaller than \(v\). Such an edge exists because \(G\) is connected,
and it is unique by minimality. Thus \(p(v)\prec v\) is uniquely determined for
every \(v\neq r\). Iterating \(p\) strictly decreases the ordering until \(r\)
is reached, so the edges \(\{v,p(v)\}\) contain no cycle and connect every
vertex to \(r\). Hence they form the claimed rooted tree.
\end{proof}
\begin{definition}
    In the setting of \Cref{thm:BHT}, define the \textbf{linking vertex}
    \[
        \lk(r)=\emptyset,\qquad \lk(v)=\max_{\prec} V(m(v)) \quad (v\neq r),
    \]
    using the convention $f(\emptyset)=+\infty$ from \Cref{sec:gwf} for the root's linking value.
    The \textbf{Basin Hierarchy Tree} (BHT) of $\gs$ with respect to $\prec$ is the pair $\tree=(T,\lk)$. The partial order induced by the rooted tree is denoted by $\preceq_T$, with \(p(v)\prec_T v\) for \(v\neq r\). The set of all BHTs of \((G,f)\), over all $\gs$-orderings, is denoted by \(\bht(G,f)\). 
    \end{definition}

\begin{proposition}
\label{thm:bht-correspondence}
For any $\tree\in\bht(G,f)$, we have
 \[
   \PD_0(G,f) \eqpd \left\{\Big[f(v), f\big(\lk(v)\big)\Big)\ \bigg|\ v\in V
   \right\}.
    \] 
\end{proposition}
\begin{proof}
For \(v\neq r\), the merge edge \(m(v)\) is exactly the edge that kills the connected component born at \(v\). Hence \(v\) contributes the interval \([f(v),\fmax(m(v)))=[f(v),f(\lk(v)))\). The root \(r\) contributes the essential interval \([f(r),\infty)=[f(r),f(\lk(r)))\). 
\end{proof}

In addition to the persistence interval indexed by each vertex, the BHT records
its parent and linking vertex. The definition of the LPF below uses these data
to determine which vertex values are raised.

\begin{algorithm}
\caption{Construction of a BHT.
During the loop, $\text{root}(v)$ denotes the root of the current tree containing
$v$. The initial value $p(v)=v$ is temporary; in the final tree only the global
minimum is its own parent.}\label{alg:bht}
\begin{algorithmic}[1]
    \Require{$\gs = ((V, E), f)$ and a $\gs$-ordering $\prec$}
    \Ensure{A BHT $\tree=(T,\lk)$}
    \State $p(v) \gets v$ for all $v\in V$ 
    \State $\lk(v) \gets \emptyset$ for all $v\in V$
    \For{$e = uv\in E$, following the $\gs$-ordering}
            \If{$\text{root}(v)\prec \text{root}(u)$}
                \State $\lk(\text{root}(u)) \gets \max_{\prec} V(e)$
                \State $p(\text{root}(u)) \gets\text{root}(v)$ 
            \ElsIf{$\text{root}(u)\prec \text{root}(v)$} 
                \State $\lk(\text{root}(v)) \gets \max_{\prec} V(e)$
                \State $p(\text{root}(v)) \gets\text{root}(u)$ 
            \EndIf
    \EndFor
\end{algorithmic}
\end{algorithm}

The following lemma gives the monotonicity of the birth levels, linking levels,
and persistence along the tree order.

\begin{lemma}\label{thm:parent-child}
    Let $\tree=(T,\lk)$ be a BHT. If \(u\preceq_T v\), then
    \begin{align}
        f(u) &\le f(v), \label{eq:bht-birth-monotone}\\
        f\big(\lk(v)\big) &\le f\big(\lk(u)\big), \label{eq:bht-link-monotone}\\
        \pers_\tree(v) &\le \pers_\tree(u), \label{eq:bht-pers-monotone}\\
        f(v) &\le f\big(\lk(u)\big), \label{eq:bht-desc-bound}
    \end{align}
    where
    \[
        \pers_\tree (v) = f(\lk(v)) - f(v), \quad v\in V .
    \]
\end{lemma}

\begin{example}
We consider the six-vertex graph and signal in \Cref{fig:bht-example}. Its BHT
is computed with respect to the unique $\gs$-ordering satisfying
$wx\prec vx$ and $vy\prec zy$. Each node $v$ represents a local minimum, and
$p(v)$ is the minimum of the component into which it merges at level
$f(\lk(v))$.

A related notion is the \emph{merge
tree}~\cite{morozov2013mergetree,Morozov-2020-merge-triplets}, whose nodes are
aligned with merge levels. The BHT instead retains the representative vertex
of each component together with $\lk(v)$ and $p(v)$. These vertex data are
needed because the LPF modifies the signal through the triples
$(v,\lk(v),p(v))$.
\begin{figure}[ht]
    \centering
    \includegraphics[width=0.85\linewidth]{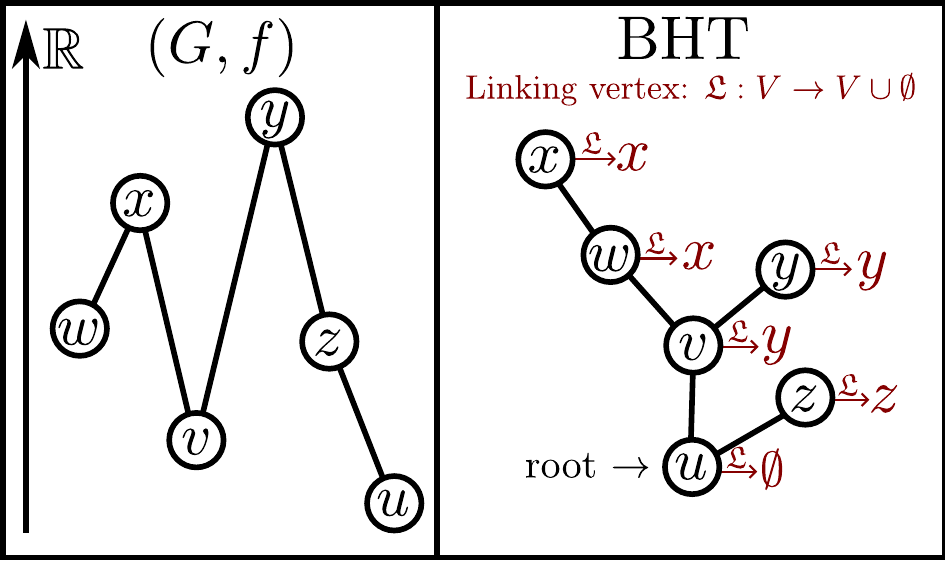}
    \caption{A signal over a graph (left) with its corresponding BHT (right). The BHT nodes are vertices of the original graph.}
    \label{fig:bht-example}
\end{figure}
\end{example}

\subsection{Low Persistence Filters}\label{sec:lpf}

The $0$-dimensional filter raises vertices in a basin of persistence below
$\e$ to the corresponding linking level.

\begin{definition}\label{def:lpf}
    For $\gs = ((V,E),f)$ a signal over a graph with $f: V\to \bR$, let $\tree = (T, \lk)\in \bht(G,f)$.
    We define the \textbf{0-Low Persistence Filter (0-LPF)} by
\begin{equation}\label{eq:lpf0}
       {}^\tree\lpf_0^\epsilon f(v) = \max \bigg( \Big\{ f(\lk(u))\ \Big|\ u\in \anc_v^\epsilon  \Big\}\cup \Big\{f(v)\Big\} \bigg),
\end{equation}
where
\begin{align}
    \anc_v^\epsilon = \left\{ u\in V\ \middle|\ \begin{aligned}\label{eq:ancestor}
          &u \preceq_T v\\
           &\pers_\tree(u) < \epsilon
       \end{aligned} \right\}
    \end{align} 
    is the set of ancestors of $v\in V$ in $\tree$ with persistence less than $\epsilon$. When it is clear from the context, we omit $\tree$ in the notation and write $\lpf_0^\epsilon$ for ${}^\tree\lpf_0^\epsilon$.
\end{definition}
The definition and \Cref{thm:parent-child} imply:
\begin{itemize}
    \item if $\pers_\tree(u)\ge\e$, then every ancestor $w\preceq_T u$ satisfies $\pers_\tree(w)\ge\pers_\tree(u)\ge\e$, so $\anc_u^\epsilon=\emptyset$ and the value is unchanged, ${}^\tree\lpf_0^\epsilon f(u)=f(u)$;
    \item if $0<\pers_\tree(u)<\e$, then $u\in\anc_u^\epsilon$, so the value is raised to at least its linking level, ${}^\tree\lpf_0^\epsilon f(u)\ge f(\lk(u))>f(u)$.
\end{itemize}

If $u$ is selected, then $f(\lk(u))$ occurs in the maximum in
\eqref{eq:lpf0} for every descendant of $u$. Thus the filter raises that
descendant set to at least the linking level of $u$.

\begin{theorem}\label{thm:lpf-0}
    Given a signal over a graph $(G, f)$ and a threshold \(\epsilon>0\), for any $\tree \in \bht(G, f)$ we have
 \begin{align}\label{eq:lpf-eq1}
    \PD_0(G,{}^\tree\lpf_0^\e f) \eqpd \PD_0(G,f)_{\geq \e},
 \end{align}
 and
 \begin{align}
 0\le {}^\tree\lpf_0^\epsilon f(v)-f(v) < \epsilon,
 \end{align}  
  for all $v \in V$. Moreover, if
 \[
    D_\epsilon(G,f)=
    \max\bigl(\{0\}\cup\{\pers(I)\mid I\in \PD_0(G,f)_{<\e}\}\bigr),
 \]
 then
 \begin{align}\label{eq:lpf-distortion}
    \|{}^\tree\lpf_0^\epsilon f-f\|_\infty = D_\epsilon(G,f)<\epsilon .
 \end{align}
\end{theorem}

For an embedded graph with faces, the filters are defined through the induced
graph.

\begin{definition}\label{def:lpf-gwf}
  For a signal over an embedded \gwf\ $\Gs = (\G,f)$ with surface embedding data \(\emb\), define:
  \begin{align}
        &\lpf_0^\e f = \Big(\lpf_0^\e \indf\Big)\Big|_V, \label{eq:lpf-gwf0}\\
        &\lpf_1^\e f = -\Big(\lpf_0^\e(-\indf)\Big)\Big|_V. \label{eq:lpf-gwf1}
  \end{align}
\end{definition}

The graph LPFs in \Cref{eq:lpf-gwf0,eq:lpf-gwf1} are applied directly to the extended-real induced signals; the Supplemental Material records the convention for hole vertices with values \(\pm\infty\). Each graph LPF uses an \emph{admissible} BHT for the corresponding induced signal. For \(\indf\), admissible means that the underlying ordering places every face vertex \(\face\) after some incident vertex \(a\in V(\face)\) with \(f(a)=\fmax(\face)\); for \(-\indf\), it means that the ordering sorts cells by their dual values and breaks ties by \emph{reversing} an admissible primal ordering, keeping vertices before edges within each level.

Combining \Cref{thm:lpf-0} and \Cref{thm:induced-graph-PH} gives the following
corollary. The distinction between the full and finite diagrams in its two
lines is essential on a surface of positive genus.

\begin{corollary}\label{thm:lpf-corollary}
 For any \(\epsilon>0\),
 \begin{align}
   \PD_0(\G, \lpf_0^\epsilon f)
   &\eqpd \PD_0(\G, f)_{\geq\e},\label{eq:lpf-corollary-0}\\
   \PD_1(\G, \lpf_1^\epsilon f)^{\mathrm{fin}}
   &\eqpd \bigl(\PD_1(\G, f)_{\geq\e}\bigr)^{\mathrm{fin}}.
   \label{eq:lpf-corollary-1}
 \end{align}
 In both cases,
 \[
   \|\lpf_n^\epsilon f - f\|_\infty < \epsilon,
   \qquad n=0,1.
 \]
 The number $r$ of essential handle intervals is fixed by the embedded support.
 If $r=0$, as in the planar case, \eqref{eq:lpf-corollary-1} is an equality of
 the complete diagrams up to zero-persistence intervals. For $r>0$, the birth
 levels of the essential handle intervals are not asserted to be preserved.
\end{corollary}

Although $\lpf_0^\e$ preserves the $0$-dimensional intervals of persistence at
least $\e$ and $\lpf_1^\e$ preserves the finite $1$-dimensional intervals of
persistence at least $\e$, each may
alter intervals in the other dimension. Consequently, neither filter alone
solves \Cref{prob:reformulated}, and even the single composition
$\lpf_1^\epsilon \lpf_0^\epsilon$ does not: as
\Cref{fig:single-composition-insufficient} shows, subthreshold intervals can
remain in $\lpf_1^\epsilon \lpf_0^\epsilon f$.

\begin{figure}[ht]
    \centering
    \includegraphics[width=0.85\linewidth]{fig/single-composition-insufficient.pdf}
    \caption{An example where $\lpf_1^\epsilon\lpf_0^\epsilon$ does not suffice. With $\epsilon=3$, ${\PD_0(\G,\lpf_1^\epsilon\lpf_0^\epsilon f)_{<\e} = \{[1,3)\} \neq \emptyset}$.}
    \label{fig:single-composition-insufficient}
\end{figure}

To address this problem, we consider an iterative application of $\lpf_0^\e$ and $\lpf_1^\e$:
\begin{equation}\label{eq:alternating-filter}
\begin{aligned}
    g_0&=f,\\
    g_{k+1}&=
    \begin{cases}
        \lpf_0^\epsilon g_k, & \text{if \(k\) is even},\\
        \lpf_1^\epsilon g_k, & \text{if \(k\) is odd},
    \end{cases}
    \qquad
    \lpf_*^{\epsilon,k}f \deff g_k .
\end{aligned}
\end{equation}
When equal values occur, each occurrence of \(\lpf_0^\e\) or \(\lpf_1^\e\)
depends on an admissible BHT choice. We use the following fixed tie-breaking
scheme: at each step, the BHT is rebuilt from the current signal, and ties are
broken by the initial ordering of the original signal. The details are given
in the Supplemental Material.
The following theorem gives finite termination and the properties of the
output.
\begin{theorem}\label{thm:lpf-all}
    For any embedded $\Gs=(\G,f)$ and threshold \(\epsilon>0\), the sequence
    \(\{\lpf_*^{\epsilon,k}f\}_{k\geq0}\), constructed with the fixed
    tie-breaking scheme above, stabilises after finitely many steps. Denote by
    \(k_o\) the smallest index such that
    \(\lpf_*^{\epsilon,k}f=\lpf_*^{\epsilon,k_o}f\) for all \(k\geq k_o\), and define
    \begin{equation}\label{eq:lpf-all}
        \lpf_*^\epsilon f \deff \lpf_*^{\epsilon,k_o} f.
    \end{equation}
    Then $\|f-\lpf_*^\epsilon f\|_\infty < \e$, ${\PD_n(\G,\lpf_*^\epsilon f)_{<\e} \eqpd \emptyset}$ for $n=0,1$, and $\lpf_*^\epsilon f \lowerpd f$.
\end{theorem}

\section{Qualitative Examples and Applications}\label{sec:experiments}

We evaluate the Low Persistence Filter using its Python implementation\footnote{\url{https://github.com/mvlier/topapprox}}.
For independent checks of persistence diagrams in array-valued examples, we use Cubical Ripser\footnote{\url{https://github.com/shizuo-kaji/CubicalRipser}}.

For reproducibility, we adopt the following conventions throughout this section. Each signal is displayed on the same value scale as the vertical axis of its persistence diagram, so that a threshold can be read directly against the diagram. The threshold is reported either as an absolute value $\e$ or, when more informative, as the percentage of finite persistence intervals it removes. Unless stated otherwise, persistence is computed in dimensions $0$ and $1$, and the alternating filter is iterated to the fixed point $\lpf_*^{\e,k_o}$ of \Cref{thm:lpf-all}; for the single-dimension examples only the relevant $\lpf_0^\e$ or $\lpf_1^\e$ is applied.

\begin{example}
We first consider a simple $1$D signal, such as a time series, represented on a line graph (\Cref{fig:line-signal-lpf-thresholds}, left).
The original signal exhibits high-frequency fluctuations. Applying the LPF with
different thresholds suppresses shallow basins (middle and right). The
corresponding persistence diagrams are shown below the signals. Each point
represents a local minimum paired with its persistence value. For \(\e=1.2\),
the large-scale trend is retained; for \(\e=2.4\), additional basins are
removed.
The unfilled boundary-side basin in the largest-threshold output corresponds to the essential $0$-dimensional interval; it has infinite persistence and is therefore not selected by any finite threshold.
\begin{figure}[!ht]
    \centering
     \includegraphics[width=0.48\textwidth]{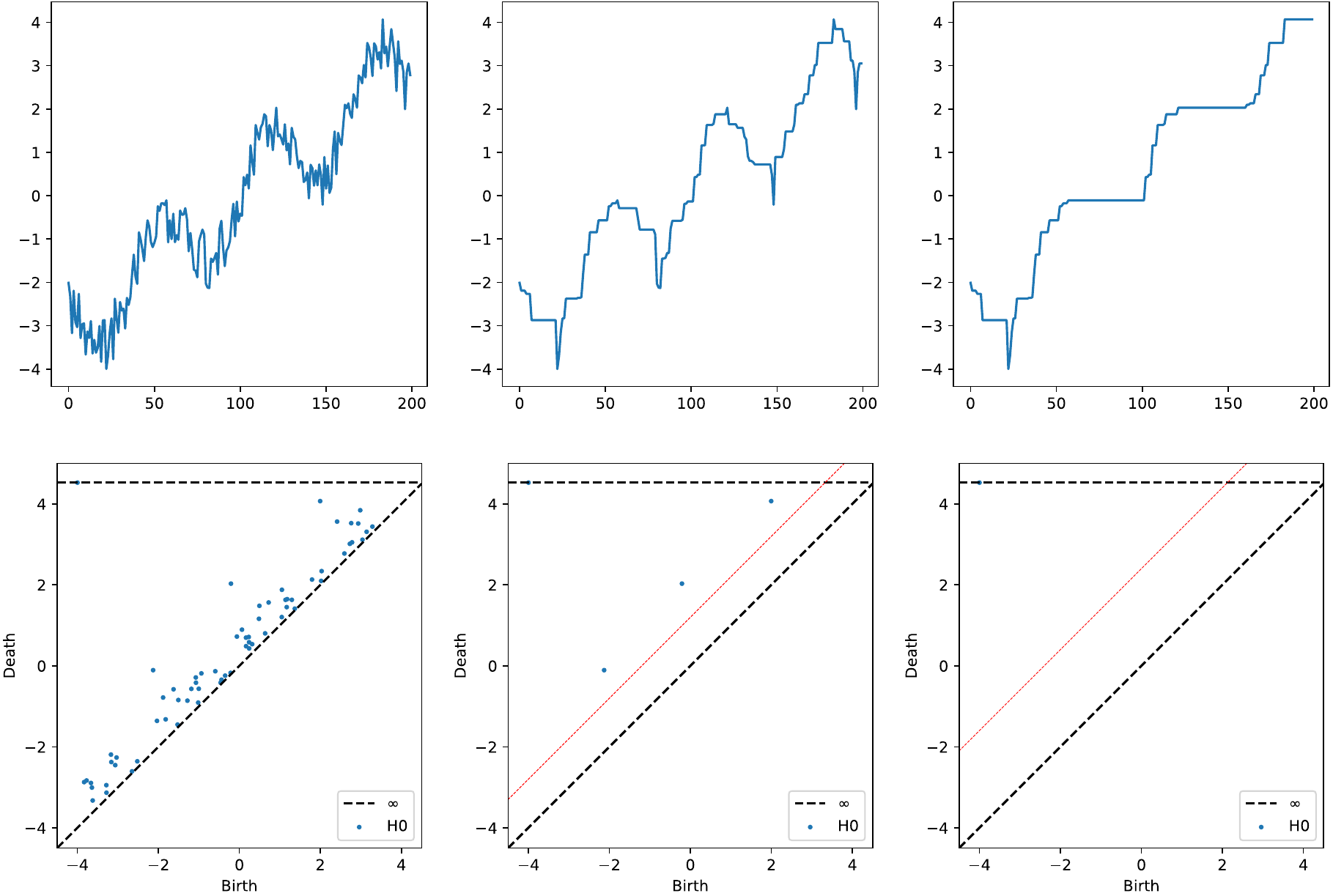}
    \caption{LPF applied to a $1$D signal on a line graph with thresholds
    \(\e=1.2\) (middle) and \(\e=2.4\) (right). The larger threshold removes
    additional basins.}
    \label{fig:line-signal-lpf-thresholds}
\end{figure}
\end{example}

\begin{example}
This example compares the persistence-based and frequency-based notions of
``low'' and ``high'' discussed in \Cref{sec:related-work}. A frequency-domain
low-pass filter retains slowly oscillating components and discards rapidly
oscillating ones, whereas the LPF retains high-persistence features and removes
low-persistence features independently of their oscillation rate. The two
criteria differ, for example, for a narrow high-persistence feature or a smooth
low-persistence feature.

We construct such a signal $f$ on a line graph with $N=1000$ vertices as the sum of three ingredients (\Cref{fig:filter-comparison}, top): two narrow triangular spikes of height $2$ that should be \emph{kept}; a low-frequency sinusoidal ripple of amplitude $0.16$ at $3$ cycles over the domain, and aperiodic broadband noise of amplitude $0.05$. Each spike is narrow, hence \emph{broadband} in frequency, yet it is a tall feature of high persistence. The ripple, conversely, is the lowest non-constant frequency present, yet its persistence (about $0.32$) is small. 

We apply the alternating LPF with threshold $\e=0.4$, iterated to
$\lpf_*^{\e,k_o}$ as in \Cref{thm:lpf-all}. This removes intervals below
$\e$ in both $f$ and the dual signal $-f$. For comparison, we apply an ideal
brick-wall Fourier low-pass filter that keeps the lowest $k$ frequency bins,
using the two cutoffs $k=10$ and $k=70$.

The outputs are shown in the middle and bottom rows of
\Cref{fig:filter-comparison}. The LPF output remains within $\e$ of the input
and retains both spike heights at $2.00$. The Fourier filter with $k=10$
reduces the spike height to $0.32$ and exhibits Gibbs ringing; with $k=70$, the
spike height is $1.68$, but $114$ local extrema remain.

The power spectrum in \Cref{fig:filter-comparison-spectrum} explains this
difference. The energy of a narrow spike is distributed over a broad frequency
range and overlaps the noise spectrum, while the ripple lies below both
cutoffs. Hence a simple frequency cutoff cannot remove the ripple and broadband
noise while retaining the complete spike spectrum.

\Cref{tab:filter-comparison} summarises the comparison through three indicators: the recovered spike height (target $2.00$), the number of local extrema (a proxy for residual oscillation; the input has $656$, the clean target $3$), and the maximum modification $\|f_{\mathrm{out}}-f\|_\infty$. The LPF preserves the spike height ($2.00$), reduces the signal to $5$ extrema, and $\|f_{\mathrm{out}}-f\|_\infty = 0.38 < \e = 0.4$, in agreement with the guarantee of \Cref{thm:lpf-all}.
The Fourier low cutoff changes the signal by $1.68$, while the high cutoff
leaves $114$ extrema. A smooth low-pass filter, such as a Gaussian or
Butterworth filter, would reduce the ringing but would still attenuate the
spikes while passing the low-frequency ripple.

\begin{figure}[!t]
  \centering
  \includegraphics[width=\columnwidth]{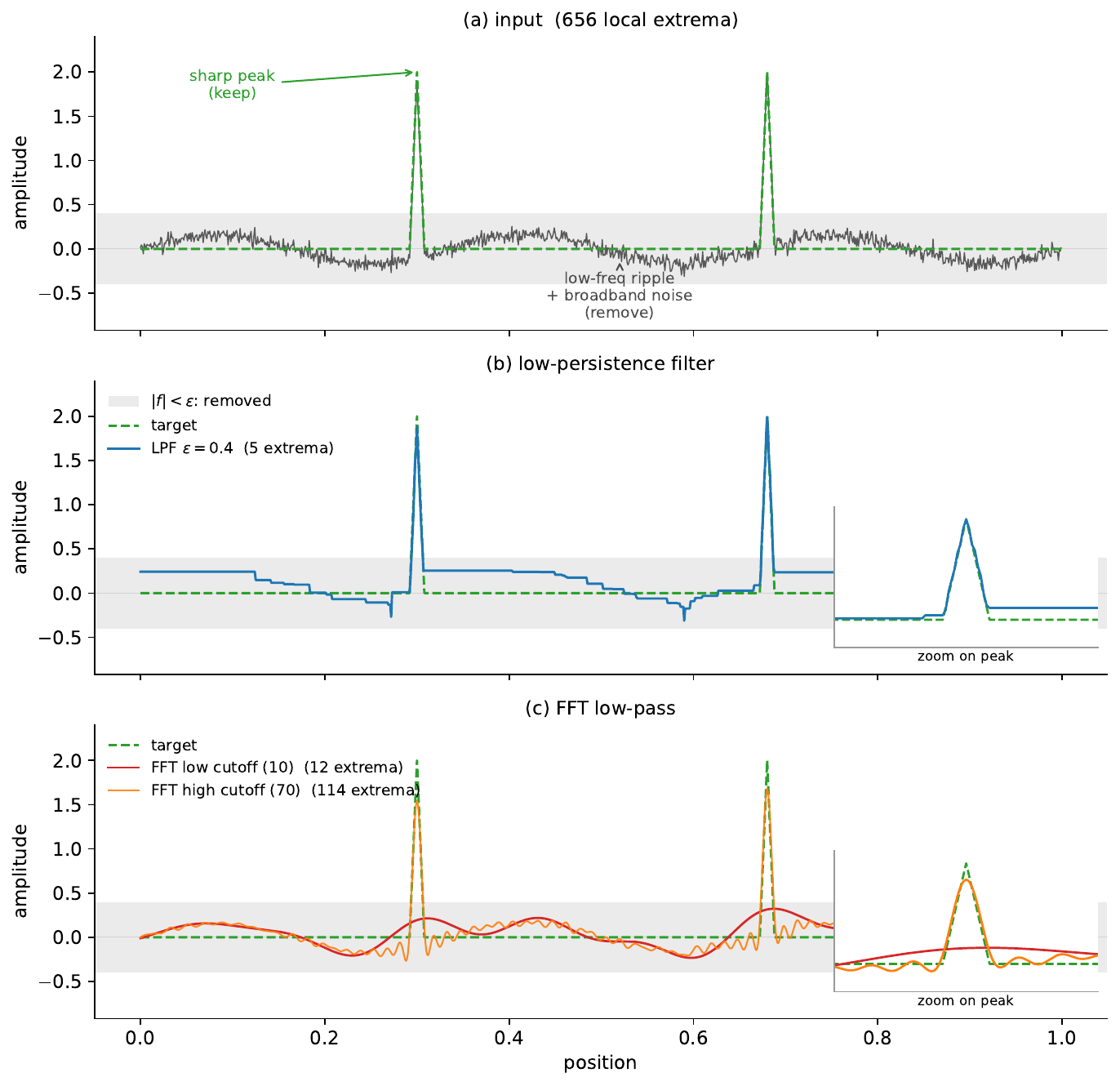}
  \caption{Persistence-based versus frequency-based filtering of a $1$D signal. \textbf{Top:} the input, two sharp spikes (target, dashed) plus a low-frequency ripple and broadband noise; the shaded band marks the $\pm\e$ subthreshold zone with $\e=0.4$. \textbf{Middle:} the LPF preserves the spikes and collapses the small features into the band. \textbf{Bottom:} a Fourier low-pass at a low cutoff ($k=10$) flattens the spikes, while a high cutoff ($k=70$) keeps them but retains noise and Gibbs ringing. Insets zoom on the right-hand spike.}
  \label{fig:filter-comparison}
\end{figure}

\begin{figure}[!t]
  \centering
  \includegraphics[width=\columnwidth]{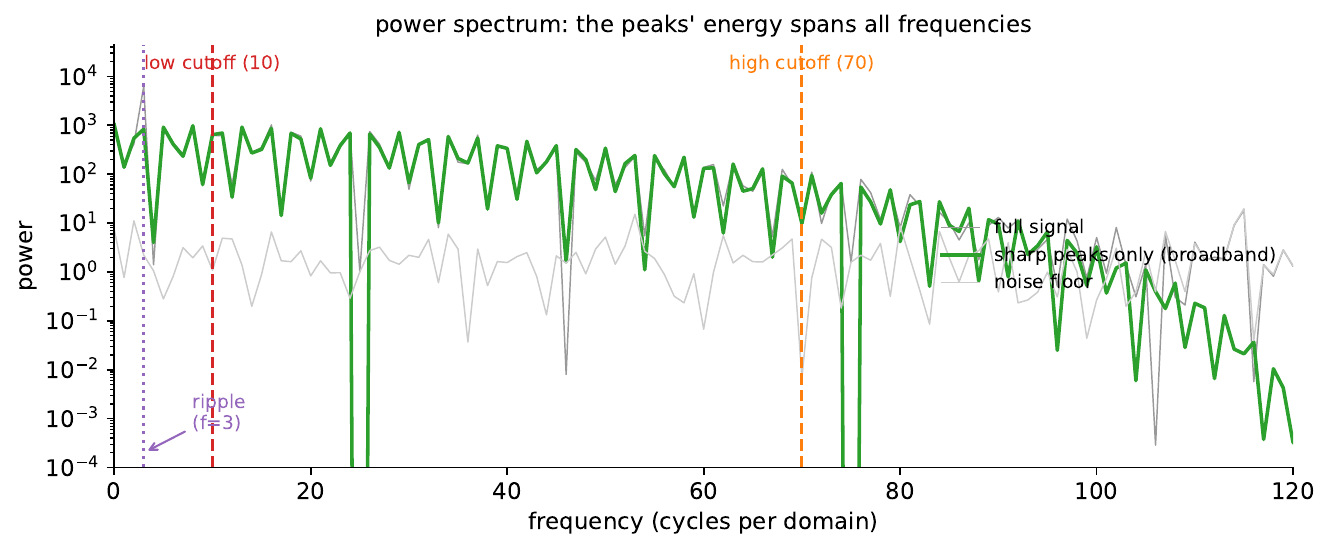}
  \caption{Power spectrum of the signal in \Cref{fig:filter-comparison}. The
  spike spectrum (green) occupies a broad frequency range and overlaps the
  noise spectrum. The low-frequency ripple (dotted line, $f=3$) lies below both
  cutoffs (dashed lines).}
  \label{fig:filter-comparison-spectrum}
\end{figure}

\begin{table}[!t]
  \centering
  \caption{Quantitative comparison for the signal of \Cref{fig:filter-comparison}. Spike height has target $2.0$; the input has $656$ local extrema and the clean target $3$. The LPF modification is provably bounded by $\e=0.4$.}
  \label{tab:filter-comparison}
  {\footnotesize
  \setlength{\tabcolsep}{4pt}
  \begin{tabular}{lccc}
    \hline
    Method & spike height & \# extrema & $\|f_{\mathrm{out}}-f\|_\infty$ \\
    \hline
    input / target        & $2.00$ & $656 / 3$ &  \\
    LPF ($\e=0.4$)        & $2.00$ & $5$       & $0.38$ \\
    Fourier low ($k=10$)  & $0.32$ & $12$      & $1.68$ \\
    Fourier high ($k=70$) & $1.68$ & $114$     & $0.33$ \\
    \hline
  \end{tabular}}
\end{table}
\end{example}

\begin{example}
A $2$D function can be viewed as a signal over a grid graph with faces.
\Cref{fig:shekel-function-lpf} shows a modified Shekel function on a
\(500\times500\) grid with many local extrema. Applying the LPF with threshold
\(\e=10^5\) removes all finite-persistence intervals, leaving only the
essential class.

\begin{figure}[ht]
  \centering
    \includegraphics[width=\subfiguresize]{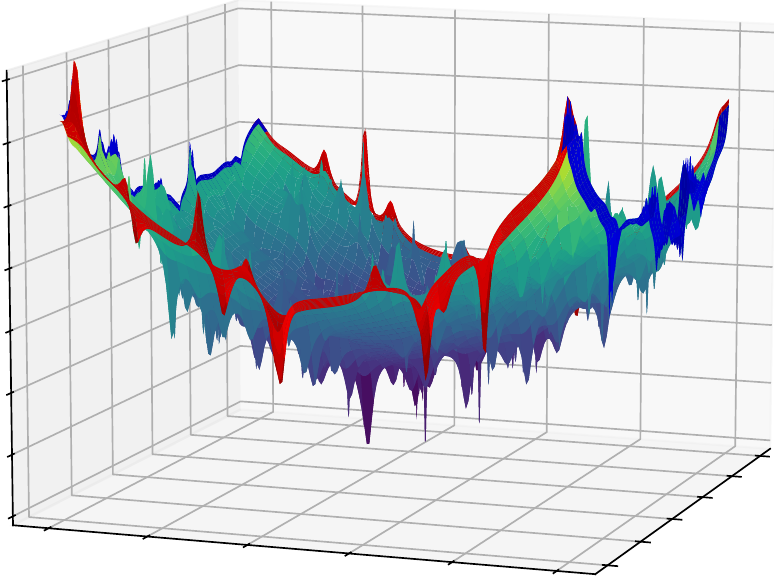}
  \hfill
    \includegraphics[width=\subfiguresize]{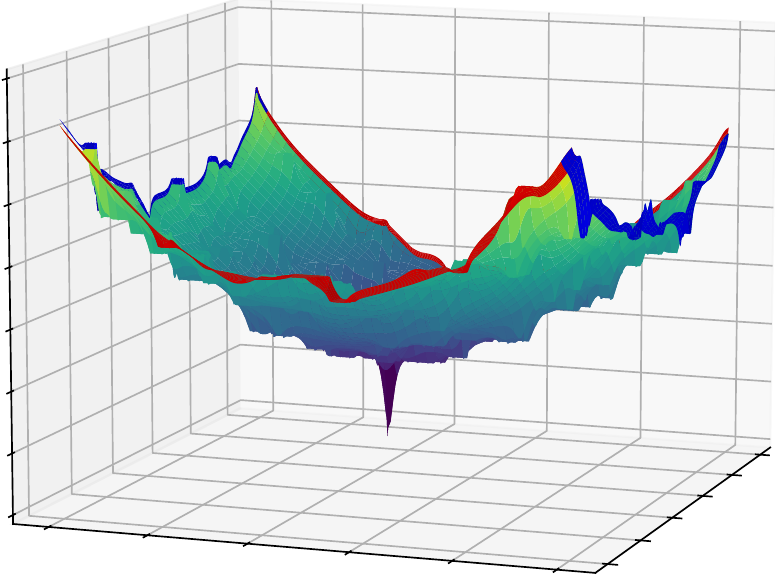}
  \caption{Modified Shekel function (left) and LPF output (right). The threshold is \(\e=10^5\), which removes all finite-persistence intervals, leaving only the global minimum (the essential class).}
  \label{fig:shekel-function-lpf}
\end{figure}
\end{example}

\begin{example}
An image can be interpreted as a signal on a grid graph with square faces,
where pixel intensities provide the signal values. \Cref{fig:pagoda-lpf} shows
the LPF applied to a natural image.

The original image and its corresponding persistence diagram are shown in the top row. Pixel intensities are normalised to the interval \([0,1]\). The blue and red dashed lines in the persistence diagram indicate the thresholds \(\e=0.0704\) and \(\e=0.3022\), which filter 90.51\% and 99.78\% of the persistence intervals, respectively. The resulting images are shown in the lower row.

Removing $90.51\%$ of the finite persistence intervals produces little visible
change, whereas removing $99.78\%$ leaves fewer image structures. Thus the
larger threshold produces a stronger simplification.

\begin{figure}[ht]
  \centering
  \begin{tabular}{@{}c@{}}
    \includegraphics[width=\subfiguresize]{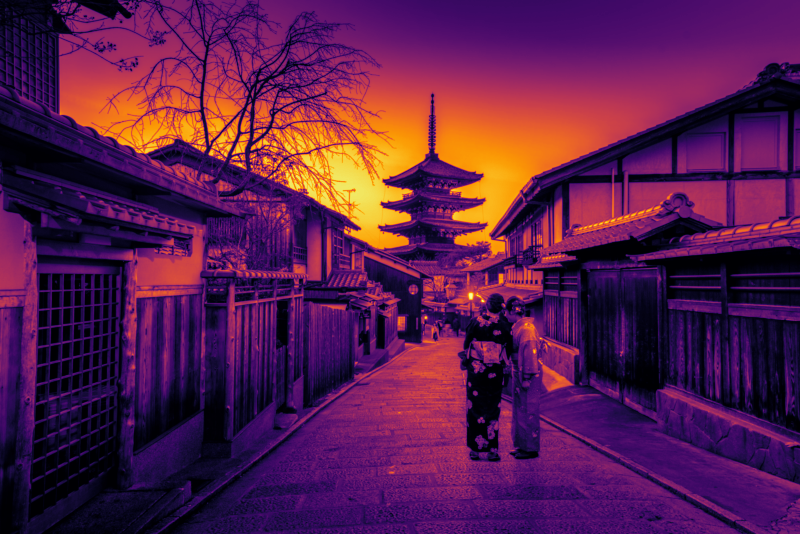} \\[\abovecaptionskip]
    \small original\\ 
  \end{tabular}
  \hfill
  \begin{tabular}{@{}c@{}}
    \includegraphics[width=\subfiguresize]{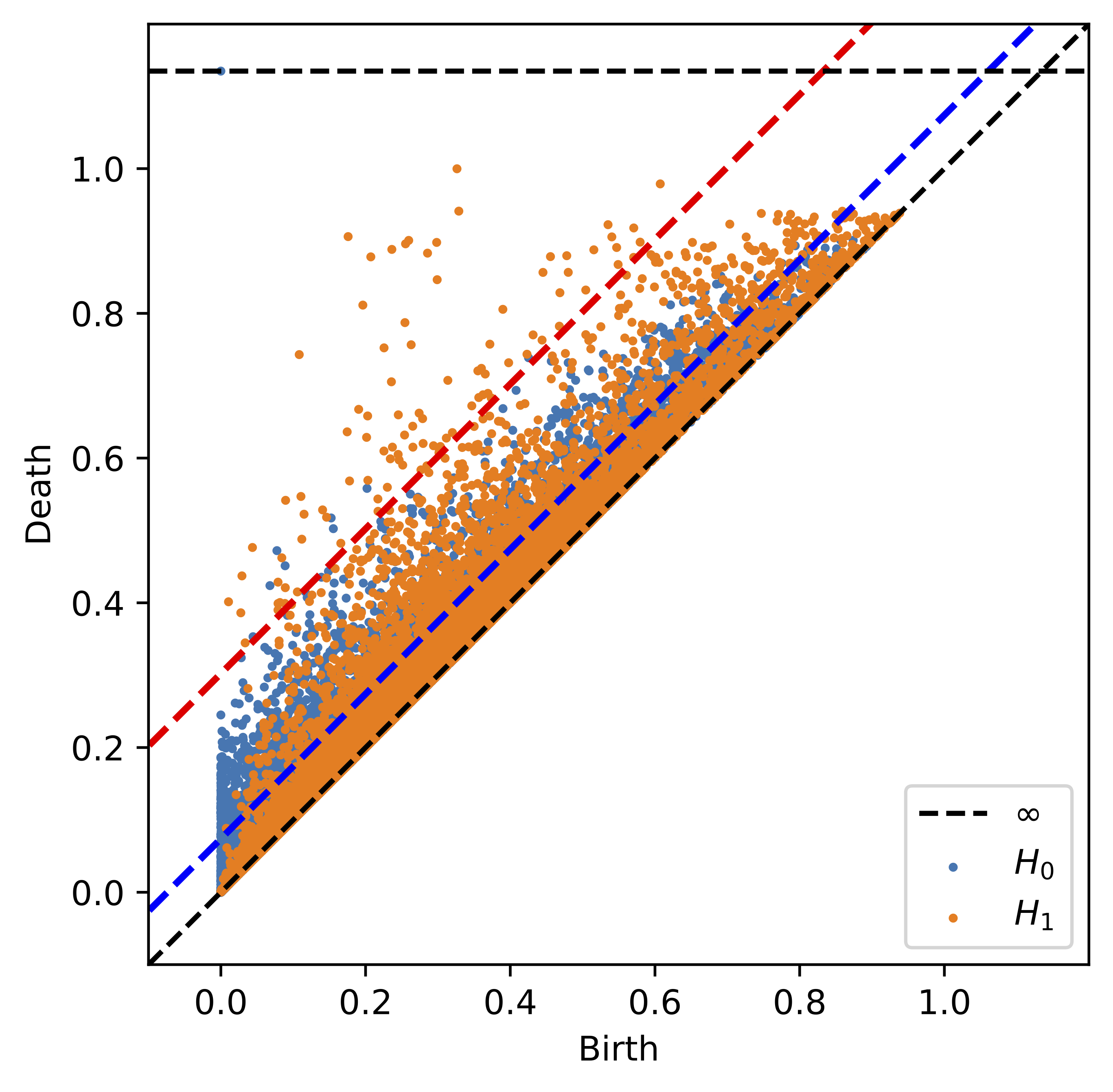}
  \end{tabular}
  \vfill
  \begin{tabular}{@{}c@{}}
    \includegraphics[width=\subfiguresize]{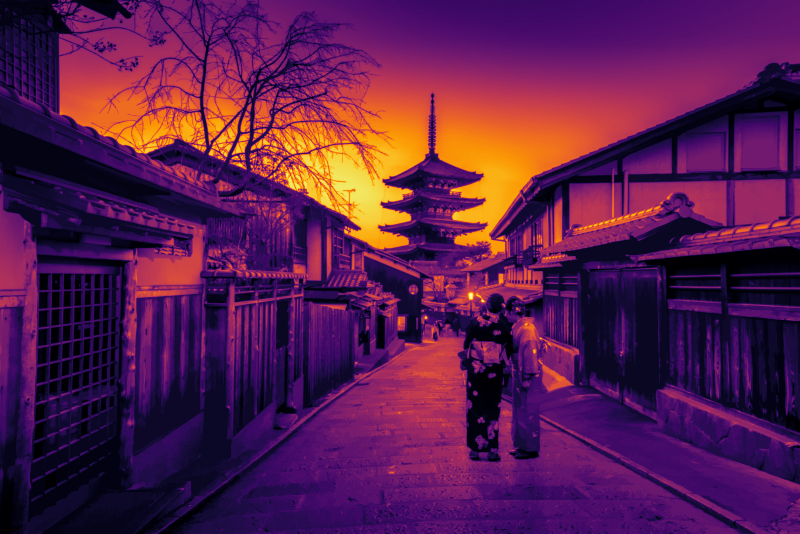} \\[\abovecaptionskip]
    \small 90.51\% filtered
  \end{tabular}
  \hfill
  \begin{tabular}{@{}c@{}}
    \includegraphics[width=\subfiguresize]{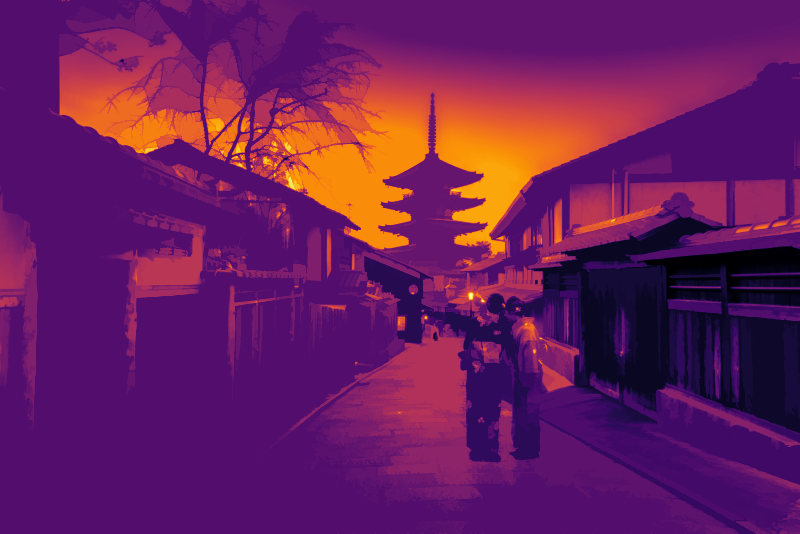} \\[\abovecaptionskip]
    \small 99.78\% filtered
  \end{tabular}
  \caption{Natural image, its persistence diagram, and LPF outputs for two filtering levels. The thresholds are \(\e=0.0704\) and \(\e=0.3022\) on the normalised intensity scale.}
  \label{fig:pagoda-lpf}
\end{figure}
\end{example}

\begin{example}
A function defined on a triangular surface mesh can be considered a signal
over a graph with triangular faces. \Cref{fig:mesh-lpf-sequence} shows the LPF
applied to such a signal.
The top row shows the original signal and its persistence diagram. 
The signal is a manifold harmonic, which is an eigenfunction of the mesh Laplacian.
In the middle row, the LPF is applied to the $0$-dimensional persistence, effectively removing darker ``dents.'' The bottom row shows the subsequent suppression of $1$-dimensional features, which removes brighter ``bumps.'' This second step acts through the dual signal \(-\indf\), so high-valued local structures in the original signal appear as low-persistence basins in the dual. In both cases, the persistence threshold is set to $\e=0.2$.
\setlength{\subfiguresize}{0.23\textwidth}
\begin{figure}[!ht]
  \centering
    \includegraphics[width=\subfiguresize]{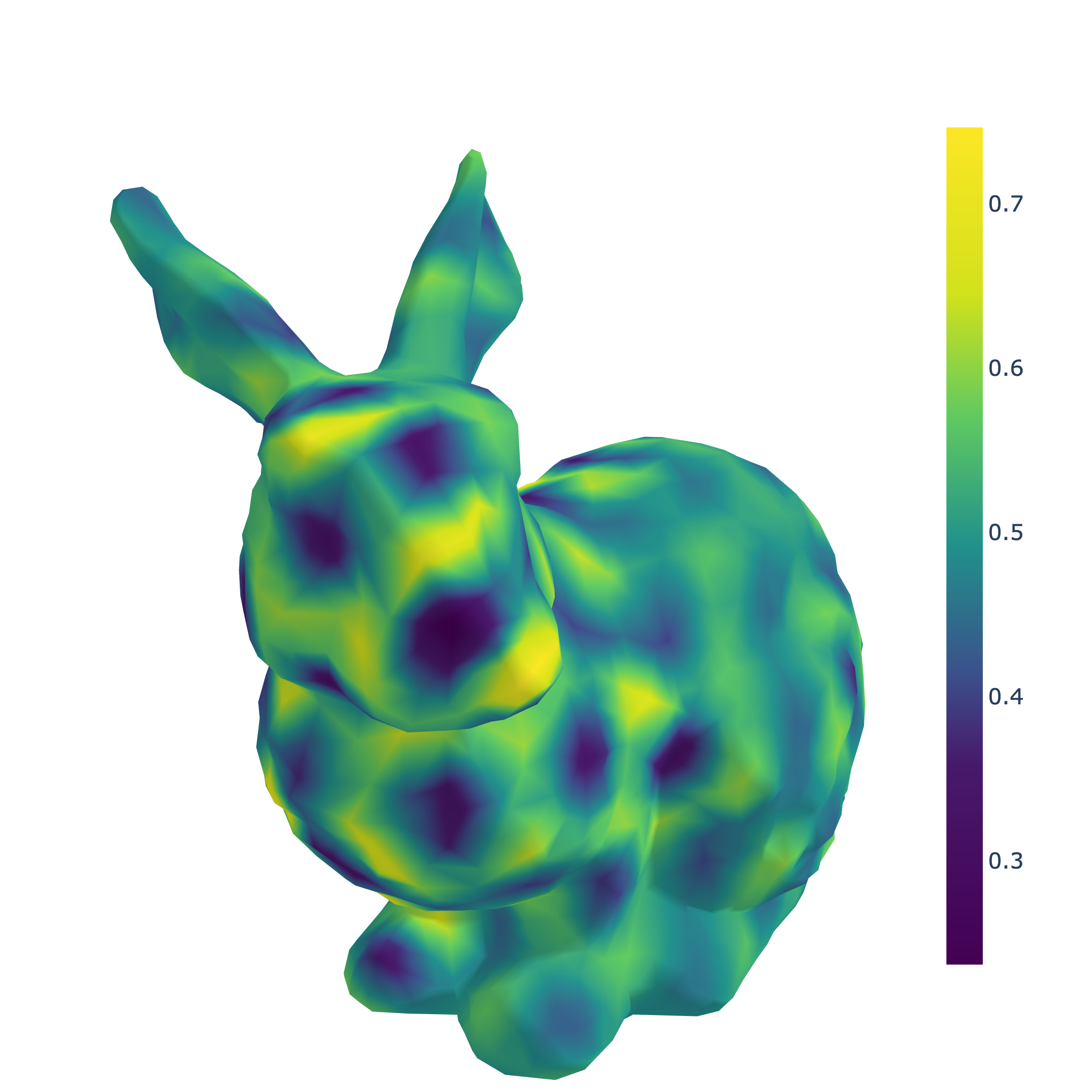}
  \hfill
    \includegraphics[width=\subfiguresize]{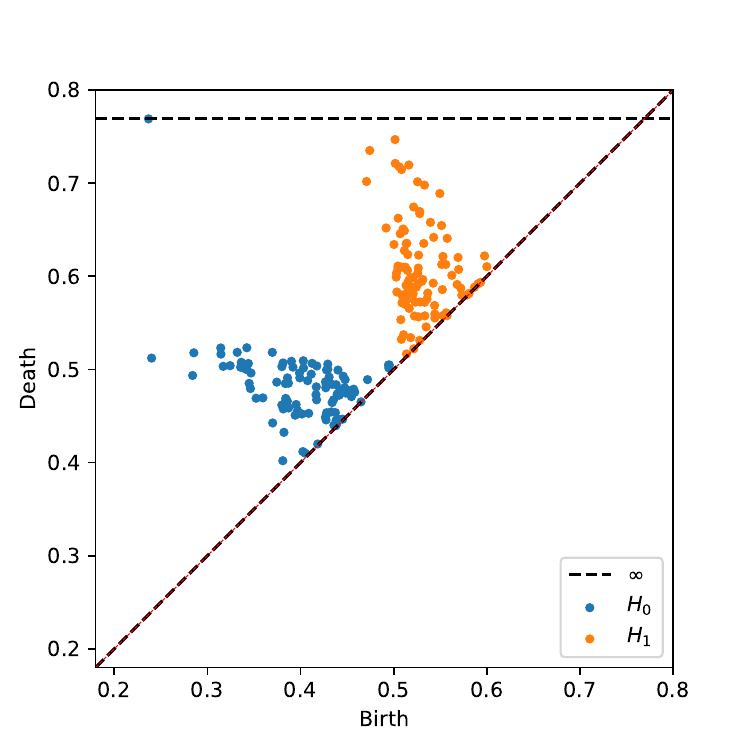}
  \vfill
    \includegraphics[width=\subfiguresize]{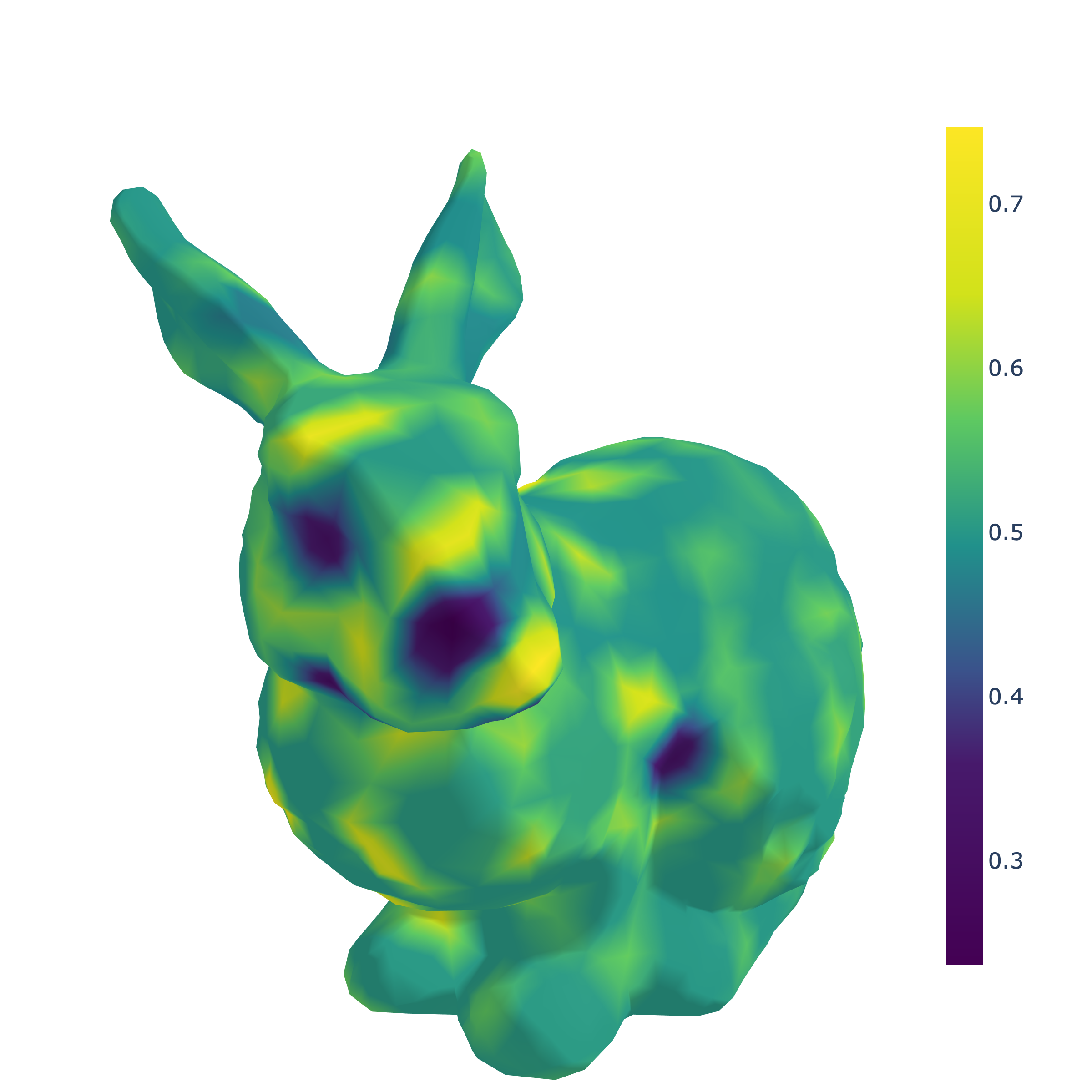}
  \hfill
    \includegraphics[width=\subfiguresize]{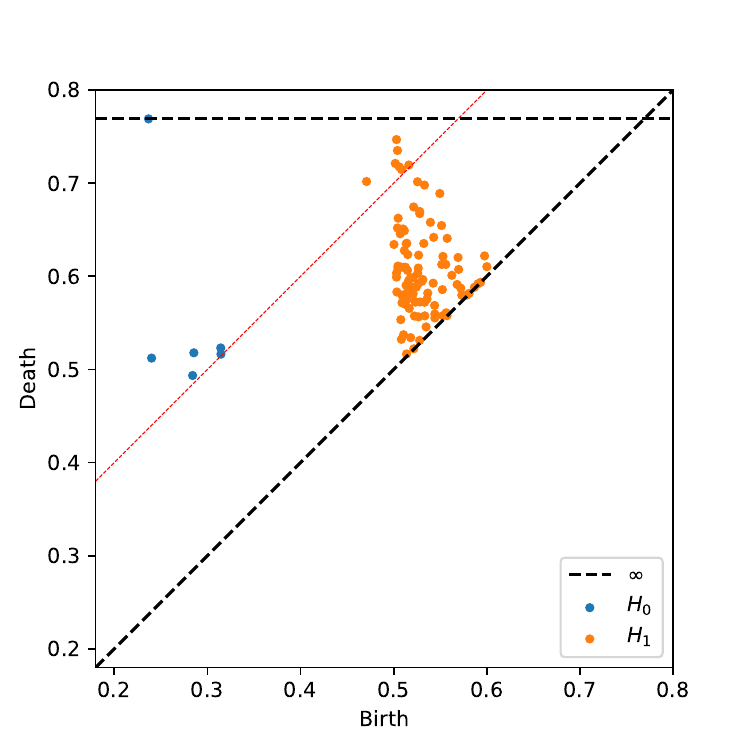}
  \vfill
    \includegraphics[width=\subfiguresize]{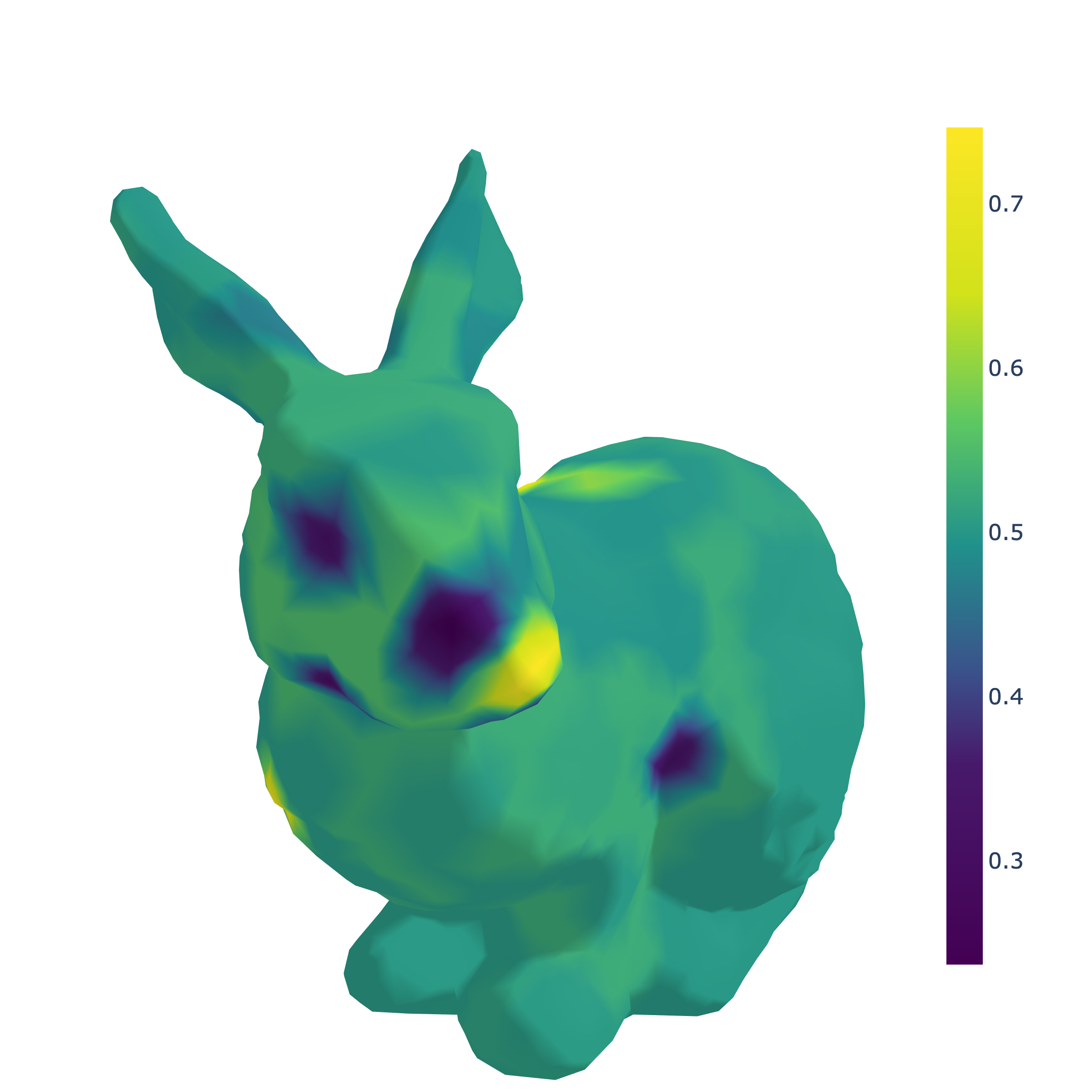}
  \hfill
    \includegraphics[width=\subfiguresize]{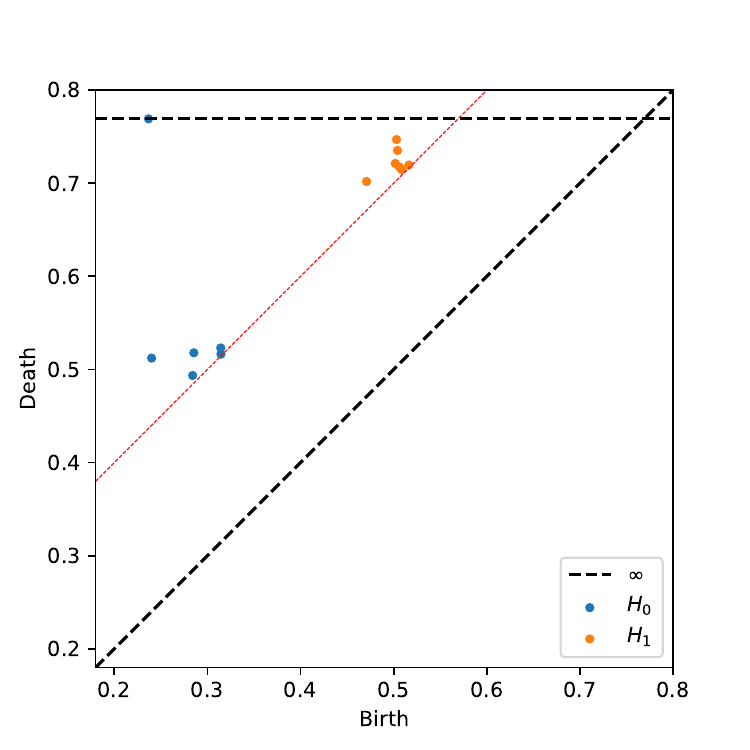}
  \caption{Signal on a triangular surface, treated as a graph with faces (top), followed by LPF outputs after suppressing low-persistence $0$-dimensional features (middle) and then low-persistence $1$-dimensional features (bottom). Here $\e=0.2$.}
  \label{fig:mesh-lpf-sequence}
\end{figure}
\end{example}


\section{Conclusion}\label{sec:conclusion}
We studied topology-aware filtering of vertex-defined graph signals under the constraint that the combinatorial support is kept fixed and only vertex values are modified.
We showed that exact simultaneous removal of low-persistence features in dimensions $0$ and $1$ is impossible in general, which motivates the relaxed formulation analysed in the paper.
The resulting Low Persistence Filter is nonlinear and suppresses subthreshold
finite-persistence artefacts while controlling the $\ell_\infty$ distortion.
Its selection criterion is an application-dependent persistence threshold,
rather than graph frequency.
On graphs with faces embedded in a surface, the method addresses both connected
components and loops through the induced-graph construction. The BHT records
the vertex-level basin relations used by the filtering algorithm.
Future work includes data-driven threshold selection, quantitative comparisons
with other filtering methods, learning tasks based on BHT representations,
applications to optimisation, and the design of high-persistence and band-pass
topological filters.

\section*{Acknowledgment}
The authors are grateful to Junyan Chu and Tomoki Uda for their valuable comments and suggestions. The authors also thank the referees for their insightful comments and careful reading of the manuscript.

\bibliographystyle{IEEEtran}
\bibliography{MyReferences}

\ifincludesupplement
  \clearpage

\begingroup
\setlist{itemsep=0.5em}
\renewcommand{\thesection}{S\arabic{section}}
\setcounter{section}{0}
\setcounter{subsection}{0}
\setcounter{equation}{0}
\setcounter{theorem}{0}
\makeatletter
\@ifundefined{theHsection}{}{\renewcommand{\theHsection}{supplement.\arabic{section}}}
\@ifundefined{theHsubsection}{}{\renewcommand{\theHsubsection}{supplement.\arabic{section}.\arabic{subsection}}}
\@ifundefined{theHequation}{}{\renewcommand{\theHequation}{supplement.\arabic{section}.\arabic{equation}}}
\@ifundefined{theHtheorem}{}{\renewcommand{\theHtheorem}{supplement.\arabic{section}.\arabic{theorem}}}
\@ifundefined{theHlemma}{}{\renewcommand{\theHlemma}{supplement.\arabic{section}.\arabic{lemma}}}
\@ifundefined{theHcorollary}{}{\renewcommand{\theHcorollary}{supplement.\arabic{section}.\arabic{corollary}}}
\@ifundefined{theHproposition}{}{\renewcommand{\theHproposition}{supplement.\arabic{section}.\arabic{proposition}}}
\@ifundefined{theHdefinition}{}{\renewcommand{\theHdefinition}{supplement.\arabic{section}.\arabic{definition}}}
\@ifundefined{theHremark}{}{\renewcommand{\theHremark}{supplement.\arabic{section}.\arabic{remark}}}
\makeatother

\section{Background on Homology and Persistent Homology}\label{app:background}


This section gives a self-contained, intuition-first account of the two
topological notions used throughout the paper---\emph{homology} and
\emph{persistent homology}---for readers who are not specialists in topology; it
is not used in the proofs and may be skipped by readers already familiar with the
subject. To keep the exposition concrete, we deliberately restrict to the cases
the paper actually needs: homological dimensions $0$ and $1$, coefficients in the
two-element field $\F=\{0,1\}$, and the sublevel-set construction applied to a
vertex-defined signal. The general theory, in arbitrary dimension and with
arbitrary coefficients, can be found in standard
references~\cite{algtop-Hatcher,TDA-Edelsbrunner,TDA-ghrist}.

\subsection{\texorpdfstring{Homology in dimensions $0$ and $1$}{Homology in dimensions 0 and 1}}\label{app:homology}

\paragraph{Intuition} Homology attaches to a space $X$ a sequence of vector
spaces $H_0(X),H_1(X),\dots$ that count its ``holes'', organised by dimension.
Only the first two are relevant here, and both have an elementary reading:
\begin{itemize}
  \item $\dim H_0(X)$ is the number of \emph{connected components} (separate
        pieces) of $X$;
  \item $\dim H_1(X)$ is the number of \emph{independent loops}
        (one-dimensional holes): closed curves in $X$ that cannot be shrunk to a
        point within $X$ and are not filled in by a two-dimensional region.
\end{itemize}
For instance, a single point has $\dim H_0=1$ and $\dim H_1=0$; a circle has
$\dim H_0=1$ and $\dim H_1=1$; two disjoint circles have $\dim H_0=2$ and
$\dim H_1=2$; a filled disk has $\dim H_1=0$, because its boundary loop is
filled. The dimensions $b_i\deff\dim_{\F} H_i(X)$ are the (mod-$2$) \emph{Betti
numbers}.

\paragraph{Algebraic definition} For the simplicial, cubical, and cell complexes
that arise from our examples ($1$D signals on line graphs, images on grids, and
triangular meshes), the Betti numbers are computed as follows. We work over the
field $\F=\{0,1\}$, in which arithmetic is taken modulo $2$. This coefficient
choice is what makes the bookkeeping elementary: since $-1=1$, orientations and
signs play no role, and a sum of cells can be identified with the \emph{set} of
cells it contains. Let $C_0,C_1,C_2$ be the $\F$-vector spaces with bases given
by the vertices $V$, the edges $E$, and the faces $F$, respectively. The
\emph{boundary maps}
\[
  C_2 \xrightarrow{\ \partial_2\ } C_1 \xrightarrow{\ \partial_1\ } C_0
\]
send each edge to the sum of its two endpoints,
\(
  \partial_1(uv)=u+v,
\)
and each face to the (mod-$2$) sum of the edges on its boundary. One checks that
$\partial_1\partial_2=0$, i.e.\ the boundary of a face is a closed loop. The
homology vector spaces are the quotients
\[
  H_0(X)=C_0/\operatorname{im}\partial_1,
  \qquad
  H_1(X)=\ker\partial_1/\operatorname{im}\partial_2 .
\]
Concretely, an element of $\ker\partial_1$---an edge set in which every vertex
meets an even number of selected edges---is a \emph{cycle}, and a cycle is
\emph{trivial} (zero in $H_1$) exactly when it is the boundary of some
combination of faces. Thus $b_0$ counts connected components and $b_1$ counts
the loops that are \emph{not} filled by faces, matching the intuition above.

\paragraph{Remark on graphs with faces} The main text
(\Cref{sec:homology}) adopts a slightly broader, geometric definition: the
homology of a \gwf\ embedded in a surface is the singular homology of its
embedded space $\G_\emb$. This is needed only to accommodate faces that are not
disk-like; on the disk-faced complexes above it agrees with the chain-level
description, and in every case the reading ``$H_0=$ components, $H_1=$ loops''
remains valid. 

\subsection{Sublevel sets and Persistent homology}\label{app:filtration}

Persistent homology turns a single signal into a one-parameter family of spaces
and tracks how their homology evolves. Let $f:V\to\R$ be the vertex-defined
signal, and recall the induced cell values $\fmax(\sigma)=\max f(V(\sigma))$ of
\Cref{sec:gwf}. For a level $t\in\R$, the \emph{sublevel set} $\G_t$ is the
subcomplex consisting of every cell $\sigma$ with $\fmax(\sigma)\le t$. As $t$
sweeps upward, these sets are nested and grow from empty to the whole complex,
\[
  \emptyset=\G_{-\infty}\subseteq\cdots\subseteq\G_s\subseteq\G_t
  \subseteq\cdots\subseteq\G_{+\infty}=\G
  \qquad(s\le t).
\]
This increasing family is the \emph{sublevel filtration}. It records what one
sees while raising a water level
across the landscape of the signal: basins (valleys around local minima) appear
and merge, and loops form and are eventually filled.

As the level $t$ increases, topological features are created and later
destroyed:
\begin{itemize}
  \item In dimension $0$, a new connected component is \emph{born} when $t$ first
        reaches a local minimum, and two components \emph{die} (merge) when $t$
        reaches the edge joining them. When two components meet, the one born at
        the lower level survives (the \emph{elder rule}).
  \item In dimension $1$, a loop is \emph{born} when $t$ closes a cycle of edges
        and \emph{dies} when $t$ reaches a face that fills it.
\end{itemize}
Each feature is summarised by the pair of levels $(b,d)$ at which it is born and
dies; its \emph{persistence} is the lifetime $d-b\ge 0$. Collecting all such
pairs in dimension $i$ yields the \emph{persistence diagram}
$\PD_i(\G,f)$, a multiset of half-open intervals $[b,d)\subset\bR$. The same data
is often drawn as points $(b,d)$ lying above the diagonal, or as a
\emph{barcode} of horizontal bars. Some features never die---for example, the
single component that remains after everything has merged, or a loop that is
never filled. These are called \emph{essential} and are recorded with $d=+\infty$
(hence infinite persistence). Throughout the paper,  diagrams keep
the essential intervals but omit the zero-persistence intervals $[a,a)$, which
carry no topological information.

\paragraph{Stability of persistence diagrams} The stability
theorem~\cite{Edelsbrunner-stability-2005} states that
if two signals differ by at most $\delta$ in the supremum norm,
$\|f-g\|_\infty\le\delta$, then their persistence diagrams differ by at most
$\delta$ in the bottleneck distance. Consequently, any feature whose
persistence exceeds $2\delta$ must be matched to a feature in the other
diagram, whereas features close to the diagonal may be matched to the
diagonal. This motivates treating low-persistence features as
\emph{topological noise}. The threshold $\e$ in the main text separates the
features to be suppressed (persistence below $\e$) from those to be preserved
(persistence at least $\e$).

\paragraph{Connection to the constructions in this paper} The $0$-dimensional
diagram $\PD_0(\G,f)$ tracks the merging of basins; the Basin Hierarchy Tree of
\Cref{sec:bht} encodes this process at the vertex level, recording
for each basin its representative minimum and the level at which it merges. The
$1$-dimensional diagram $\PD_1(\G,f)$ is obtained, via \Cref{thm:induced-graph-PH},
from the $0$-dimensional persistence of a dual signal, so the same construction
handles loops. The per-dimension Low Persistence Filters delete the targeted
finite intervals below~$\e$ while preserving the longer-lived intervals in the
corresponding dimension. The alternating filter removes all subthreshold finite
intervals in dimensions $0$ and $1$ under the monotone simplification relation
of \Cref{def:lower-pd}; essential intervals are never selected, and their number
is fixed by the support.

\section{Technical Conventions}\label{app:technical-conventions}

\begin{remark}[Extended-real convention for hole vertices]\label{rem:extended-real-hole}
The induced graph assigns the value \(+\infty\) to hole vertices, and the dual induced signal therefore assigns them the value \(-\infty\). We use these extended-real values directly. The orderings place \(+\infty\)-valued cells after all finite cells and \(-\infty\)-valued cells before all finite cells, with admissible tie-breaking inside the infinite level. Persistence is computed in the extended order: an interval with exactly one endpoint at \(\pm\infty\) has infinite persistence, while a formal interval \([\infty,\infty)\) or \([-\infty,-\infty)\) has zero persistence. Consequently, hole vertices contribute only essential, infinite-persistence, or zero-persistence formal pairs. They never contribute a positive finite interval below a finite threshold \(\epsilon\), and any zero-persistence update at an infinite level changes no finite vertex value. Thus the LPF on \(\indf\) or \(-\indf\) is interpreted directly in the extended real line, and the filtered values on \(V\) are unaffected by the infinite bookkeeping vertices.
\end{remark}

\begin{remark}[Extension of orderings to the induced graph]\label{rem:ordering-extension}
We record an extension of orderings from the original \gwf\ to its induced graph.
Let $\prec$ be a $\Gs$-ordering. We extend $\prec$, originally defined on $V\cup E\cup F$, to a $(G[\emb],\indf)$-ordering on
\[
V'\cup E' = V\cup F\cup \hole \cup E\cup E_F\cup E_{\hole}.
\]
The elements of $F$ become vertices in $G[\emb]$, and their relative order with respect to $V$ and $E$ is preserved. For each face $\face\in F$, all incident edges in $E_F$ are placed immediately after $\face$. Finally, all vertices in $\hole$ are placed after that, followed by all edges in $E_{\hole}$. Within $\hole$, $E_F$, and $E_{\hole}$, the internal order is arbitrary. This extension is a valid ordering: an $E_F$ edge $(\face,v)$ has $\indf$-value $\fmax(\face)$ because $f(v)\le\fmax(\face)$, so placing it immediately after $\face$ preserves value-monotonicity, and its endpoints precede it since $v\prec e\prec\face$ for some $e\in\face$ by \eqref{eq:g-ord-edge} and \eqref{eq:g-ord-face2}; the cells at $+\infty$ are covered by the convention of \Cref{rem:extended-real-hole}.
\end{remark}

\begin{remark}[Compatible orderings]\label{rem:compatible-ordering}
Given a $\gs$-ordering $\prec$, there exists a unique $(G,g)$-ordering for another signal $g$ obtained by sorting first by the cell values $g^\uparrow$ and then by the reference order $\prec$ to break ties:
\begin{equation}\label{eq:induced-order}
\begin{aligned}
\sigma_1 \prec' \sigma_2
\iff{}&
\bigl(g^\uparrow(\sigma_1)<g^\uparrow(\sigma_2)\bigr)\\
&\text{or }
\bigl(g^\uparrow(\sigma_1)=g^\uparrow(\sigma_2)\text{ and }\sigma_1\prec\sigma_2\bigr).
\end{aligned}
\end{equation}
We call $\prec'$ the $(G,g)$-ordering \textbf{compatible} with $\prec$.
\end{remark}

\begin{remark}[Dependence on admissible BHT choices]\label{rem:lpf-choice-supp}
The graph filter ${}^\tree\lpf_0^\epsilon$ depends on the chosen admissible BHT $\tree$ (\Cref{sec:lpf}), equivalently on the tie-breaking used to order cells of equal value. \Cref{thm:lpf-0} holds for every such admissible BHT.

For the alternating filter in \Cref{thm:lpf-all}, fix admissible initial
orderings \(\prec_0\) for \(\ind f\) and \(\prec_1\) for \(-\ind f\)
(\Cref{sec:lpf}). A deterministic implementation may order the primal cells
lexicographically by value, then cell dimension, then a fixed vertex or cell
index, and order the dual cells by dual value, breaking ties by the reverse of
the primal rule and placing vertices before edges. At every step \(k\), rebuild
the two BHTs for the current signal \(g_k\), breaking ties by \(\prec_0\) for
the \(0\)-LPF and by \(\prec_1\) for the \(1\)-LPF. Because
\Cref{thm:lpf-0} and \Cref{lem:stability} hold for every admissible BHT, this
fixed tie-breaking scheme yields the reduction in persistence required in the
proof of convergence.

For this scheme, \Cref{prop:fixed-tie-monotonicity} shows that the ancestor
relations of each BHT are monotone with respect to the initial signal. This
property is used to prove the \(\ell_\infty\) bound in \Cref{thm:lpf-all}.
\end{remark}

\section{Proofs}\label{app:proofs}

This section collects the proofs deferred from the main paper.
They are ordered by dependency: first the induced-graph representation, then
the graph LPF, and finally the interaction of the primal and dual filters used
by the alternating construction.

Throughout the proof of \Cref{thm:induced-graph-PH}, fix a signal \(\Gs=(\G,f)\) over a \gwf\ embedded in a connected closed surface \(\Sigma\) with surface embedding \(\emb=(\gamma,\phi)\). For \(t\in\R\), write \(V_t,E_t,F_t\) for the sets of cells with \(\fmax\le t\) and define the \textbf{embedded sublevel space}
\[
    X_t \deff \gamma\bigl((V_t,E_t)\bigr)\ \cup \bigcup_{\face\in F_t}\cl{\phi(\face)}\ \subseteq\ \Sigma .
\]
If \(\fmax(\face)\le t\), then \(\partial\phi(\face)=\bigcup_{e\in\face}\gamma(e)\subseteq\gamma(E_t)\),
so \(X_t\cap\gamma\bigl((V,E)\bigr)=\gamma\bigl((V_t,E_t)\bigr)\).
Therefore, \((\gamma,\phi)\) restricts to a surface embedding of the sublevel \gwf\ \(\G_t\) with embedded space \(X_t\); the persistent homology of \(\Gs\) is therefore the persistence module of the nested family \(\{X_t\}_t\), and \(X_t=\G_\emb\) for \(t\ge\max f\). The space \(X_t\) is the union of the compact connected \emph{pieces} \(\gamma(v)\) (\(v\in V_t\)), \(\gamma(e)\) (\(e\in E_t\)), and \(\cl{\phi(\face)}\) (\(\face\in F_t\)); its complement \(\Sigma\setminus X_t\) is the union of the connected \emph{complementary pieces}: the points \(\gamma(v)\) with \(f(v)>t\), the arcs \(\gamma(e)\setminus\gamma(V_t)\) with \(\fmax(e)>t\), the regions \(\phi(\face)\) with \(\fmax(\face)>t\), and the holes. 

\begin{lemma}\label{lem:complement-incidence}
Let \(t\in\R\), let \(G[\emb]_t=\{\indf\le t\}\) be the sublevel graph, and let \(K_t\) be the subgraph of \(G[\emb]\) induced by the vertices \(\sigma\) with \(\indf(\sigma)>t\). There are bijections
between the connected components
\[
    \pi_0\bigl(G[\emb]_t\bigr)\cong\pi_0(X_t),
    \qquad
    \pi_0(K_t)\cong\pi_0(\Sigma\setminus X_t),
\]
natural in \(t\): for \(s\le t\) they commute with the component maps induced by the inclusions \(X_s\subseteq X_t\), \(G[\emb]_s\subseteq G[\emb]_t\), \(\Sigma\setminus X_t\subseteq\Sigma\setminus X_s\), and \(K_t\subseteq K_s\). In particular, \(t<\min f\) gives \(K_t=G[\emb]\) and \(\Sigma\setminus X_t=\Sigma\), so \(G[\emb]\) is connected.
\end{lemma}

\begin{proof}
We use the following elementary fact. Let \(U\) be a finite union of connected pieces \(P_1,\dots,P_m\), and call two pieces \emph{touching} when \(P_i\cap\cl{P_j}\neq\emptyset\) or \(\cl{P_i}\cap P_j\neq\emptyset\). Then the components of \(U\) are exactly the unions of the touching-connectivity classes: touching pieces lie in one component, since adjoining to \(P_j\) a point of \(P_i\cap\cl{P_j}\) preserves connectedness, while distinct classes are separated in \(U\) because closure commutes with finite unions.

\emph{Sublevel side.} The pieces of \(X_t\) correspond to the cells of \(\G_t\). Since \(\gamma\) is an embedding and \(\cl{\phi(\face)}\) meets the embedded graph exactly in \(\partial\phi(\face)=\bigcup_{e\in\face}\gamma(e)\), two pieces touch exactly when the corresponding cells share an incident vertex: an edge with its endpoints, two edges with a common endpoint, a face with the endpoints \(V(\face)\) of its boundary edges. Each such shared vertex \(w\) satisfies \(f(w)\le t\), and in \(G[\emb]_t\) it joins the same pair of cells: directly for edges, and through the \(E_F\) edge \((\face,w)\), present at the level \(\fmax(\face)\le t\), for faces. Hence the touching classes of the pieces of \(X_t\) coincide with the components of \(G[\emb]_t\), which gives the first bijection.

\emph{Complement side.} Each complementary piece contains or touches the image of a vertex of \(K_t\): a point piece \(\gamma(v)\) is such an image; an arc piece of an unentered edge \(e\) contains the image of an unentered endpoint of \(e\) (at least one endpoint satisfies \(f>t\)); a region or hole corresponds to its own vertex of \(K_t\). Conversely, every edge of \(K_t\) is realised by touching pieces: for \(e\in E\) with both endpoints unentered, the arc piece of \(e\) joins them; for \((\face,v)\in E_F\) and \((h,v)\in E_{\hole}\), the point \(\gamma(v)\) lies on the boundary of the region. It remains to check that every touching between complementary pieces is mirrored in \(K_t\). Any face \(\face\) flanking an unentered edge \(e\) contains \(e\) in its boundary list, so \(\fmax(\face)\ge\fmax(e)>t\) and \(\face\) is itself a complementary region; an arc piece of \(e\) touches only its unentered endpoints and the two flanking regions, and each unentered endpoint \(v\) of \(e\) satisfies \(v\in V(\face)\) (or \(\gamma(v)\in\partial h\)), so the flanking regions and the endpoints are pairwise joined in \(K_t\) through \(v\). A region or hole touches a point piece \(\gamma(v)\) exactly when \(\gamma(v)\) lies on its boundary, which for a face means \(v\in V(\face)\) and for a hole is the defining relation of \(E_{\hole}\); two open regions never touch each other. Finally, entered vertices and edges lie in \(X_t\), so no touching passes through them, matching the fact that \(K_t\) is an induced subgraph. Hence the touching classes of the complementary pieces coincide with the components of \(K_t\).

Naturality holds by construction: for \(s\le t\), every piece at parameter \(t\) is contained in a piece at parameter \(s\) associated with the same or an incident cell, and the inclusions \(G[\emb]_s\subseteq G[\emb]_t\), \(K_t\subseteq K_s\) match this containment cell by cell. The final assertion follows because \(\Sigma\) is connected.
\end{proof}

\subsection*{Proof of \Cref{thm:induced-graph-PH}}

Let $\Gs = ((V,E,F),f)$ be a signal over a \gwf\ embedded in a connected closed surface $\Sigma$, and let \(\emb=(\gamma,\phi)\) be a surface embedding for \(\G\). We keep the notation of \Cref{lem:complement-incidence}; in particular, $X_t$ denotes the embedded sublevel space, whose persistence module realises the persistent homology of $\Gs$.

\emph{The diagram \(\PD_0\).}
For every \(t\in\R\), \Cref{lem:complement-incidence} identifies \(\pi_0(X_t)\) with \(\pi_0\bigl(G[\emb]_t\bigr)\) naturally in \(t\), so the two \(0\)-dimensional persistence modules are isomorphic over \(\R\). Isomorphic pointwise finite-dimensional modules have equal persistence diagrams, so the two diagrams can differ only in features invisible over the real line: zero-persistence intervals, which \(\eqpd\) discards, and the distinction between a class that dies exactly at \(+\infty\)---a merge through hole vertices and \(E_{\hole}\) edges, which exists in \(G[\emb]\) but not in the \gwf---and an essential class; both are recorded as the same interval \([b,+\infty)\). Hence
\[
    \PD_0(\G, f) \eqpd \PD_0(G[\emb], \indf).
\]

\medskip
Now we analyse \(\PD_1\). For \(t\in\R\), the space \(X_t\subset\Sigma\) is compact and
locally contractible. Write \(M=\{H_1(X_t)\}_{t}\) for the \(1\)-dimensional persistence module of the sublevel filtration. For any pointwise finite-dimensional persistence module \(N\), write \(\PD(N)\) for its persistence diagram; thus \(\PD(M)=\PD_1(\G,f)\).

\emph{Step 1:} Since \(\Sigma\) is closed and we work over \(\F\), Alexander--Lefschetz duality for the compact, locally contractible subset $X_t\subseteq\Sigma$ gives $H_1(\Sigma,\Sigma\setminus X_t)\cong H^1(X_t)$~\cite[Theorem~3.44]{algtop-Hatcher}; combining it with the long exact sequence of the pair $(\Sigma,\Sigma\setminus X_t)$, with $H_1(\Sigma)\cong H^1(\Sigma)$ (Poincar\'e duality), and with $\widetilde H_0(\Sigma)=0$ yields, for each \(t\), an isomorphism
\begin{equation}\label{eq:surface-duality}
    \widetilde H_0(\Sigma\setminus X_t)\;\cong\;\operatorname{coker}\bigl(H^1(\Sigma)\to H^1(X_t)\bigr)\;\cong\;\bigl(\ker\iota_t\bigr)^{\!*},
\end{equation}
where \(\iota_t\colon H_1(X_t)\to H_1(\Sigma)\) is induced by the inclusion \(X_t\hookrightarrow\Sigma\), the last identification uses field coefficients, and $(-)^*$ denotes the linear dual. The isomorphism \eqref{eq:surface-duality} is natural in the following sense: for $s\le t$, the map on the left induced by the inclusion $\Sigma\setminus X_t\subseteq\Sigma\setminus X_s$ corresponds on the right to the linear dual of the map $\ker\iota_s\to\ker\iota_t$ obtained by restricting $H_1(X_s)\to H_1(X_t)$.
Geometrically, a complementary region is bounded by a cycle of \(X_t\) that bounds in \(\Sigma\); over \(\F\) a cycle separates \(\Sigma\) if and only if it is null-homologous in \(\Sigma\), so \eqref{eq:surface-duality} records exactly the \emph{separating} cycles of \(X_t\). When \(\Sigma=S^2\) we have \(H_1(\Sigma)=0\), hence \(\iota_t=0\) and \(\ker\iota_t=H_1(X_t)\), and \eqref{eq:surface-duality} is the reduced Alexander duality \(H_1(X_t)\cong\widetilde H_0(S^2\setminus X_t)\) of the planar case~\cite[Theorem~3.43]{algtop-Hatcher}.

\emph{Step 2:} By \Cref{lem:complement-incidence}, the components of \(\Sigma\setminus X_t\) are identified, naturally in $t$, with those of the subgraph \(K_t\) of \(G[\emb]\) induced by the vertices with \(\indf>t\). Passing from the decreasing superlevel parameter \(t\) to the increasing parameter \(s=-t\) identifies the reduced \(0\)-dimensional persistence of this complementary filtration with \(\widetilde\PD_0(G[\emb],-\indf)\), under \([a,b)\leftrightarrow[-b,-a)\); here the strict superlevel sets $\{\indf>t\}$ and the non-strict dual sublevel sets $\{-\indf\le s\}$ produce the same persistence diagram.
Together with Step~1, 
the submodule \(K=\{\ker\iota_t\}_{t}\subseteq M\) therefore satisfies
\begin{equation}\label{eq:K-is-dual}
\begin{aligned}
    \PD(K)
        &\eqpd \widetilde\PD_0(G[\emb],-\indf),\\
    [a,b) &\leftrightarrow [-b,-a).
\end{aligned}
\end{equation}

\emph{Step 3:} The maps \(\iota_t\) form a morphism from \(M\) to the constant persistence module with value \(H_1(\Sigma)\), with kernel \(K\) and image module \(Q\deff\{\operatorname{im}\iota_t\}_t\cong M/K\). Since $Q$ is a submodule of a constant module, all its structure maps are injective. We use the following elementary fact.

\emph{Persistence-diagram additivity.} Let $0\to K\to M\to Q\to 0$ be a short exact sequence of pointwise finite-dimensional persistence modules such that every structure map of $Q$ is injective. Then $\PD(M)=\PD(K)\sqcup\PD(Q)$. Indeed, for $s\le t$ write $M_{s,t}$ for the structure map of $M$. If $x\in\ker M_{s,t}$, then the image of $x$ in $Q_s$ dies under the injective map $Q_{s,t}$, so it vanishes and $x\in K_s$; hence $\ker M_{s,t}=\ker K_{s,t}$. Therefore
\begin{align*}
    \operatorname{rank} M_{s,t}
    &= \dim M_s-\dim\ker K_{s,t}\\
    &= \operatorname{rank} K_{s,t}+\dim Q_s
    = \operatorname{rank} K_{s,t}+\operatorname{rank} Q_{s,t},
\end{align*}
using $\dim M_s=\dim K_s+\dim Q_s$ and the injectivity of $Q_{s,t}$. Since the persistence diagram of a pointwise finite-dimensional module is determined linearly by its rank function (the number of intervals containing $[s,t]$ equals $\operatorname{rank} M_{s,t}$, and inclusion--exclusion over the endpoints recovers the multiplicities), additivity of ranks gives additivity of persistence diagrams.

Because all structure maps of $Q$ are injective, $\PD(Q)$ consists only of half-infinite intervals; the classes it detects are the handle classes that eventually appear in \(\operatorname{im}\iota_\infty=\operatorname{im}\bigl(H_1(\G_\emb)\to H_1(\Sigma)\bigr)\), since $X_t=\G_\emb$ for $t\ge\max f$.
Thus the persistence diagram of \(M\) splits as the persistence diagram of \(K\) together with these essential intervals:
\[
\begin{aligned}
    \PD(M)
        &= \PD(K)\ \sqcup\ \mathcal{E}_\Sigma,\\
    |\mathcal{E}_\Sigma|
        &= r
        = \dim_{\F}\operatorname{im}\bigl(H_1(\G_\emb)\to H_1(\Sigma)\bigr)
        \le b_1(\Sigma),
\end{aligned}
\]
the \(r\) intervals of \(\mathcal{E}_\Sigma\) being essential. With \eqref{eq:K-is-dual} and \(\PD(M)=\PD_1(\G,f)\), this is precisely \eqref{eq:pd1}; in particular the reversal restricts to a bijection on the finite-persistence intervals of the two diagrams. When \(\hole=\emptyset\), \(\G_\emb=\Sigma\) and \(\iota_\infty=\mathrm{id}\), so \(r=b_1(\Sigma)\); when \(\Sigma=S^2\), \(r=0\). \qed

\subsection*{Proof of \Cref{thm:parent-child}}

Let $\tree=(T,\lk)$ be the BHT of $(G,f)$ with respect to $\prec$, and let \(u\preceq_T v\).
If \(v\) is the root, then \(u=v\) and the assertions are immediate. Assume \(v\) is not the root. It is enough to prove \eqref{eq:bht-link-monotone} when \(u=p(v)\), because the general case follows by iterating along the ancestor chain.

If $p(v)$ is the root of $\tree$, then $f(\lk(p(v))) = \infty$, and the inequality
$f(\lk(v)) \leq f(\lk(p(v)))$ is trivial. Assume therefore that $p(v)$ is not the root. Let \(e=m(v)\) and \(e'=m(p(v))\). Then
\[
    f(\lk(v)) = \fmax(e), \qquad f(\lk(p(v))) = \fmax(e'),
\]
by the definition of the linking vertex. Moreover,
\[
    v = \min C(v,G_{\prec e}) \succ \min C(v,G_{\preceq e}) = p(v),
\]
so
\[
    \min C(p(v),G_{\preceq e})=p(v).
\]
Thus the component of \(p(v)\) has not merged into a smaller minimum by the time \(e\) appears. Since \(e'\) is the first edge at which that happens for \(p(v)\), we have \(e\prec e'\). Therefore,
\[
    f\big(\lk(v)\big) = \fmax(e) \leq \fmax(e') = f\big(\lk(p(v))\big).
\]

Now let \(u\preceq_T v\). Since \(u\) is an ancestor of \(v\), we have \(u\preceq v\) in the $\gs$-ordering, hence
\[
    f(u)\le f(v).
\]
Repeated application of the parent inequality gives
\[
    f\big(\lk(v)\big) \leq f\big(\lk(u)\big).
\]
Combining these two inequalities, we obtain
\[
    \pers_\tree(v) = f\big(\lk(v)\big) - f(v)
    \leq f\big(\lk(u)\big) - f(u)
    = \pers_\tree(u).
\]

Finally, \(v\preceq m(v)\) in the $\gs$-ordering, so \(f(v)\le\fmax(m(v))=f(\lk(v))\); combined with \eqref{eq:bht-link-monotone}, this gives \(f(v)\le f(\lk(v))\le f(\lk(u))\), which is \eqref{eq:bht-desc-bound}.
\qed

\subsection*{Auxiliary properties of the BHT}

\begin{lemma}\label{lem:bht-technical}
Let $\tree=(T,\lk)$ be a BHT.
\begin{enumerate}
    \item \label{bht-tech-prefix-min} At any prefix of the ordering, the
    minimum of a connected component is a BHT ancestor of every vertex in that
    component.

    \item \label{bht-tech-desc-rel} Let \(\alpha\prec m(v)\). If a component
    \(C\) of $G_{\preceq\alpha}$ contains a descendant of \(v\), then every
    member of \(C\) is a descendant of \(v\). In particular, if
    \(t<f(\lk(v))\), \(v\prec_Tu\), and
    \(C(w,G_t)=C(u,G_t)\), then \(v\preceq_Tw\).

    \item \label{bht-tech-vlink-path} For any non-root vertex \(v\), there is
    a path from \(v\) to \(\lk(v)\) whose vertex values lie in
    \([f(v),f(\lk(v))]\).
\end{enumerate}
\end{lemma}

\begin{proof}
During the BHT construction, the representative of a component changes from a
vertex to its parent precisely when that component merges into an older one.
Thus the representative at every prefix is obtained by iterating the parent
map from each vertex in the component. This proves
\ref{bht-tech-prefix-min}; it also shows that any two ancestors of one vertex
are comparable.

For \ref{bht-tech-desc-rel}, let \(u\in C\) descend from \(v\), and let \(r\)
be the minimum of \(C\). By \ref{bht-tech-prefix-min},
\(r\preceq_Tw\) for every \(w\in C\), in particular
\(r\preceq_Tu\). The ancestors \(r\) and \(v\) of \(u\) are comparable. If
\(r\prec_Tv\), then the basin of \(v\) has already merged by the stage
\(\alpha\), contrary to \(\alpha\prec m(v)\). Hence
\(v\preceq_Tr\preceq_Tw\) for every \(w\in C\). The level statement follows
from \(t<f(\lk(v))=\fmax(m(v))\).

For \ref{bht-tech-vlink-path}, put \(e=m(v)\). If \(\lk(v)\) lies in
\(C(v,G_{\prec e})\), join it to \(v\) inside that component. Otherwise join
\(v\) to the endpoint of \(e\) in the component and append \(e\). In either
case, minimality of \(v\) gives the lower bound on the path, while every vertex
precedes \(e\), whose value is \(f(\lk(v))\), giving the upper bound.
\end{proof}

\subsection*{Auxiliary lemmas for the graph LPF}

We shall use the following two consequences of the BHT construction.

\begin{lemma}\label{lem:lpf-basin-decomposition}
Let $\tree=(T,\lk)$ be a BHT of $(G,f)$ and
\(
S_\e=\{v\in V\mid\pers_\tree(v)<\e\}.
\)
Let $R_\e$ consist of the vertices of $S_\e$ having no selected proper
ancestor. For $a\in R_\e$, put
\[
    D_a=\{u\in V\mid a\preceq_Tu\},
    \qquad c_a=f(\lk(a)).
\]
Then the sets $D_a$ are pairwise disjoint and, for
$h={}^\tree\lpf_0^\e f$,
\begin{equation}\label{eq:lpf-basin-form}
    h(u)=
    \begin{cases}
        c_a, & u\in D_a\text{ for some }a\in R_\e,\\
        f(u), & u\notin\bigcup_{a\in R_\e}D_a.
    \end{cases}
\end{equation}
Moreover, $D_a$ is the vertex set of
$C(a,G_{\prec m(a)})$.
\end{lemma}

\begin{proof}
The ancestors of any vertex form a chain. Hence, if $u$ has a selected
ancestor, it has a unique one $a\in R_\e$ closest to the root; this also proves
that the sets $D_a$ are disjoint. Every other selected ancestor $v$ of $u$
satisfies $a\preceq_Tv$, so
$f(\lk(v))\le f(\lk(a))$ by \eqref{eq:bht-link-monotone}, while
$f(u)\le f(\lk(a))$ by \eqref{eq:bht-desc-bound}. Formula
\eqref{eq:lpf0} therefore gives $h(u)=c_a$. If $u$ has no selected ancestor,
the same formula gives $h(u)=f(u)$.

Before $m(a)$, every descendant basin of $a$ has already merged into the
component represented by $a$, whereas no vertex outside that descendant set
can belong to a component with minimum $a$; the latter assertion follows from
\Cref{lem:bht-technical}, part~\ref{bht-tech-prefix-min}. Hence that component has
vertex set $D_a$.
\end{proof}

\begin{lemma}\label{lem:lowered-component-boundary}
Let $\tree$ be a BHT of $(G,-f)$ and set
\[
    f'=-{}^\tree\lpf_0^\e(-f),
    \qquad L_t=\{u\in V\mid f'(u)=t<f(u)\}.
\]
Every connected component of $L_t$ is joined by an edge of $f'$-value $t$ to
a vertex $o\notin L_t$ with $f'(o)\le t$. If $f'(o)=t$, then $f'(o)=f(o)$.
\end{lemma}

\begin{proof}
Apply \Cref{lem:lpf-basin-decomposition} to $(G,-f)$. Each $a\in R_\e$
determines a descendant set on which $f'$ is equal to
$c=f(\lk(a))$. Let
\[
    B_a=C\bigl(a,(G,-f)_{\prec m(a)}\bigr)
\]
be its component just before the dual merge. The vertices of $B_a$ are exactly
the descendants already merged into this basin; they satisfy $f\ge c$, and
those with $f>c$ are lowered to $c$, whereas those with $f=c$ are unchanged.

If an edge joins $b\in B_a$ with $f(b)>c$ to $o\notin B_a$, it cannot precede
$m(a)$ in the dual ordering. Its dual value is therefore at least $-c$, which,
because $f(b)>c$, forces $f(o)\le c$. Thus $f'(o)\le c$, and equality implies
$f'(o)=f(o)=c$. Since $B_a$ is connected, each component of the vertices
lowered to $c$ meets either an unchanged vertex of $B_a$ at level $c$ or such
an edge leaving $B_a$. In both cases the connecting edge has $f'$-value $c$,
which proves the claim.
\end{proof}

\begin{proposition}\label{thm:BHT-stability}
Let $\tree$ be the BHT of $(G,f)$ with respect to $\prec$, put
$h={}^\tree\lpf_0^\e f$, and let $\prec'$ be the compatible $(G,h)$-ordering
of \eqref{eq:induced-order}. Then $\tree$ is also the BHT of $(G,h)$ with
respect to $\prec'$. Moreover, for every $v\in V$,
\begin{equation}\label{eq:bht-filtered-endpoints}
\bigl(h(v),h(\lk(v))\bigr)=
\begin{cases}
\bigl(f(v),f(\lk(v))\bigr), & \pers_\tree(v)\ge\e,\\
\bigl(h(v),h(v)\bigr), & \pers_\tree(v)<\e.
\end{cases}
\end{equation}
\end{proposition}

\subsection*{Proof of \Cref{thm:BHT-stability}}

Use the notation $R_\e,D_a,c_a$ of
\Cref{lem:lpf-basin-decomposition}, put
$D=\bigcup_{a\in R_\e}D_a$, and call the vertices of $D$ \emph{raised}.
Thus $h=c_a$ on $D_a$ and $h=f$ outside $D$.

Fix a non-root vertex $w$, and write
\[
    e=m(w),\qquad F=\fmax(e)=f(\lk(w)),
\]
and
\[
    A=C(w,G_{\prec e}),\qquad
    B=C(p(w),G_{\prec e}).
\]
These are the two components joined by $e$, with minima $w$ and $p(w)$.
By \Cref{lem:bht-technical}, $A$ consists of descendants of $w$, while
$p(w)$ is an ancestor of every vertex of $B$.

We first determine the new level of $e$:
\begin{equation}\label{eq:bht-stability-edge-level}
    h^\uparrow(e)=
    \begin{cases}
        F, & w\notin D,\\
        c_a, & w\in D_a.
    \end{cases}
\end{equation}
We use the following observation: if a vertex $z\in D_b$ ($b\in R_\e$) lies
in $A$ or $B$, and that component contains some vertex $q\notin D$, then
$m(b)\prec e$, and hence
\[
    c_b=\fmax(m(b))\le\fmax(e)=F.
\]
Indeed, if $e\preceq m(b)$, then the component of $z$ in $G_{\prec e}$
contains the descendant $z$ of $b$ at a stage strictly before $m(b)$, so
every member of that component---in particular $q$---is a descendant of
$b$ by \Cref{lem:bht-technical}, part~\ref{bht-tech-desc-rel}; this gives
$q\in D_b\subseteq D$, a contradiction.

Suppose first that $w\notin D$; then $p(w)\notin D$, because a selected
ancestor of $p(w)$ would also be a selected ancestor of $w$. The
observation, with $q=w$ on the $A$-side and $q=p(w)$ on the $B$-side, shows
that every raised endpoint of $e$ is raised by at most $F$, while the old
linking endpoint has original value $F$; hence $h^\uparrow(e)=F$.

Now let $w\in D_a$. If $w\ne a$, then
$a\preceq_Tp(w)$, so both $A$ and $B$ lie in $D_a$; no other $D_b$ can meet
them because distinct members of $R_\e$ are incomparable. Hence
both endpoints of $e$ have value $c_a$. If $w=a$, then $F=c_a$ and
$p(w)\notin D$; every vertex of $A$ lies in $D_a$ at level $c_a=F$, and the
observation with $q=p(w)$ bounds every raising level on the $B$-side by
$F$. This proves \eqref{eq:bht-stability-edge-level}.

We next show that $A$ and $B$ are still the two components immediately before
$e$ in $\prec'$. Every cell of $A\cup B$ has new value at most
$h^\uparrow(e)$ by the preceding bounds, and a tie is resolved by $\prec$, in
which it precedes $e$. Conversely, let $e''\succ e$ have an endpoint in
$A\cup B$. Then $\fmax(e'')\ge F$. If $w\notin D$ or $w=a$, this gives
$h^\uparrow(e'')\ge F=h^\uparrow(e)$. If
$w\in D_a\setminus\{a\}$, the endpoint inside $A\cup B\subseteq D_a$ already
has value $c_a=h^\uparrow(e)$. At equality, the reference order still places
$e$ before $e''$. Thus no additional edge enters either component before $e$.

The minima are also unchanged. On $A$, either all vertices have the common
value $c_a$, or $w$ is unraised and
$h(x)\ge f(x)\ge f(w)=h(w)$. The same argument applies to $B$: it lies in the
same raised basin when $w\in D_a\setminus\{a\}$, and otherwise $p(w)$ is
unraised. Ties are resolved by $\prec$, so the minima remain $w$ and $p(w)$,
with $p(w)\prec'w$: either both have the same new value, both are unraised,
or $w=a$ and
$h(p(w))=f(p(w))\le f(w)\le c_a=h(w)$. Consequently, $e$ is still the merge
edge of $w$ and its parent is still $p(w)$.

It remains to check the linking endpoint. Let $x\prec y$ be the endpoints of
$e$, so $y=\lk(w)$. If $w\in D_a\setminus\{a\}$, both endpoints have value
$c_a$ and their order is unchanged. Otherwise
$h^\uparrow(e)=F=f(y)$, so $h(y)=F$ and the compatible tie-breaking again
places $x$ before $y$. Hence the linking map is unchanged, and $\tree$ remains
the BHT.

Finally, a vertex lies in $D$ exactly when it has a selected ancestor, which,
by \eqref{eq:bht-pers-monotone}, is equivalent to
$\pers_\tree(v)<\e$. Equation \eqref{eq:bht-stability-edge-level} and the
preservation of the linking endpoint now give
\eqref{eq:bht-filtered-endpoints}; the essential case is given by the root.
\qed

\subsection*{Proof of \Cref{thm:lpf-0}}

Set \(h={}^\tree\lpf_0^\e f\). By \Cref{thm:BHT-stability}, the same BHT
\(\tree=(T,\lk)\) computes both persistence diagrams. Equation
\eqref{eq:bht-filtered-endpoints} shows that the interval indexed by \(v\) is
unchanged when \(\pers_\tree(v)\ge\e\) and collapses to a zero-persistence
interval when \(\pers_\tree(v)<\e\). Together with
\Cref{thm:bht-correspondence}, this gives
\[
    \PD_0(G,h)\eqpd\PD_0(G,f)_{\ge\e}.
\]

It remains to prove the sharp norm bound. By \eqref{eq:lpf0} and
\eqref{eq:bht-desc-bound}, \(h(v)\ge f(v)\). If \(h(v)>f(v)\), then the
maximum in \eqref{eq:lpf0} is attained at a selected ancestor
\(u\preceq_Tv\), and
\[
\begin{aligned}
    0<h(v)-f(v)
      &=f(\lk(u))-f(v)\\
      &\le f(\lk(u))-f(u)
       =\pers_\tree(u)\\
      &\le D_\e(G,f)<\e,
\end{aligned}
\]
where \eqref{eq:bht-birth-monotone} gives \(f(u)\le f(v)\), and the BHT
correspondence identifies \(\pers_\tree(u)\) with an interval of the original
diagram. Hence \(\|h-f\|_\infty\le D_\e(G,f)\).

If \(D_\e(G,f)>0\), choose \(u\) whose BHT interval has persistence
\(D_\e(G,f)\). Since \(u\) is selected and belongs to its own ancestor set,
\[
    h(u)-f(u)\ge f(\lk(u))-f(u)=D_\e(G,f).
\]
Thus equality holds. If \(D_\e(G,f)=0\), every selected linking level equals
its birth level, so \eqref{eq:lpf0} gives \(h=f\). This proves
\eqref{eq:lpf-distortion}. \qed

\begin{lemma}\label{lem:induced-lpf-consistency}
Let \(q\) be a signal over an embedded \gwf\ and apply the graph LPF on
\(G[\emb]\).
\begin{enumerate}
    \item If \(h={}^\tree\lpf_0^\e\ind{q}\) and \(g=h|_V\), then
    \(h=\ind{g}\) on \(V\cup F\).
    \item If \(h={}^\tree\lpf_0^\e(-\ind{q})\) and \(g=-h|_V\), then
    \(h=-\ind{g}\) on \(V\cup F\).
\end{enumerate}
\end{lemma}

\subsection*{Proof of \Cref{lem:induced-lpf-consistency}}

We first note the following consequence of \Cref{lem:bht-technical}. If
\(u\preceq_Tx\) and an edge \((x,y)\) enters strictly below
\(\eta(\lk(u))\), then its component contains a descendant of \(u\) before
\(m(u)\). By \Cref{lem:bht-technical}, part~\ref{bht-tech-desc-rel},
\(u\preceq_Ty\). Thus every LPF level contributed to \(x\) also contributes
to \(y\).

First put \(\eta=\ind q\) and \(h={}^\tree\lpf_0^\e\eta\). For a face
\(\face\), choose \(a\in V(\face)\) with
\(\eta(a)=\eta(\face)=q^\uparrow(\face)\), as guaranteed by admissibility.
We prove
\[
    h(\face)=\max_{v\in V(\face)}h(v).
\]
Every level contributing to \(h(\face)\) is either its original value, which
is bounded by \(h(a)\), or a linking level \(L\). If \(L=\eta(\face)\), the
same bound applies; if \(L>\eta(\face)\), the edge \((a,\face)\) enters below
\(L\), so it follows that \(h(a)\ge L\). Hence
\(h(\face)\le\max_{v\in V(\face)}h(v)\).

Conversely, fix \(v\in V(\face)\). Its original level is at most
\(\eta(\face)\le h(\face)\). A linking level \(L\le\eta(\face)\) has the same
bound, while for \(L>\eta(\face)\) the edge \((v,\face)\) enters below \(L\)
because \(\eta(v)\le\eta(\face)\). It follows that \(h(\face)\ge L\).
Thus every \(h(v)\le h(\face)\), proving the first assertion.

For the dual signal \(\eta=-\ind q\), we have
\(\eta(\face)=\min_{v\in V(\face)}\eta(v)\). Choose
\(a\in V(\face)\) attaining this minimum. For any incident \(v\), an original
or linking level \(L\) contributing to \(h(\face)\) is already bounded by
\(h(v)\) when \(L\le\eta(v)\). If \(L>\eta(v)\), the edge \((v,\face)\)
enters below \(L\), so \(h(v)\ge L\) by the preceding observation. Therefore
\(h(\face)\le\min_{v\in V(\face)}h(v)\).

Finally, every level contributing to \(h(a)\) is either
\(\eta(a)=\eta(\face)\), or a linking level \(L>\eta(\face)\); in the latter
case \((a,\face)\) enters below \(L\), and the preceding observation gives
\(h(\face)\ge L\). Hence \(h(a)\le h(\face)\), and
\[
    h(\face)=\min_{v\in V(\face)}h(v).
\]
Since \(g=-h|_V\), this is \(h(\face)=-g^\uparrow(\face)\).

The assertions are immediate on \(V\). Hole vertices remain at
\(\pm\infty\) under the convention of \Cref{rem:extended-real-hole}; they
create no positive finite subthreshold interval and do not affect the finite
face-vertex argument above. \qed

\subsection*{Auxiliary proposition for \Cref{lem:stability}}

\begin{proposition}\label{prop:BHT-dual-decrease}
    Let \(\epsilon>0\), let $\gs=((V,E),f)$ be a signal over a connected graph \(G=(V,E)\) with $f:V\to\bR$, let $\tree_0=(T_0,\lk_0)$ be the BHT with respect to a $\gs$-ordering $\prec_0$, and let $\tree_1=(T_1,\lk_1)$ be the BHT with respect to a $(G,-f)$-ordering $\prec_1$. Put
    \[
        f' \deff -\,{}^{\tree_1\!}\lpf_0^\e(-f),
    \]
    let $\prec_0'$ be the $(G,f')$-ordering compatible with $\prec_0$ as in \eqref{eq:induced-order}, and let
    \[
        \tree_0' = (T_0', \lk_0') \in \bht\bigl(G, f'\bigr)
    \]
    be the BHT of $(G,f')$ with respect to $\prec_0'$. Then:
    \begin{enumerate}[label=(\alph*), ref=(\alph*)]
        \item \label{prop-dd-pers} $\pers_{\tree_0'}(u) \leq \pers_{\tree_0}(u)$ for all $u\in V$, and the roots of $T_0'$ and $T_0$ coincide;
        \item \label{prop-dd-order} every ancestor relation of $T_0'$ respects the reference ordering: if $w\preceq_{T_0'}u$ with $w\neq u$, then $w\prec_0 u$. In particular, $w\preceq_{T_0'}u$ implies $f(w)\leq f(u)$, and more generally $\eta(w)\le\eta(u)$ for any reference signal $\eta$ that is monotone with respect to $\prec_0$.
    \end{enumerate}
\end{proposition}

\begin{proof}
By \Cref{lem:lpf-basin-decomposition} applied to $(G,-f)$, the signal
\[
    f'=-{}^{\tree_1}\lpf_0^\e(-f)
\]
satisfies $f'\le f$. Call $u$ \emph{lowered} when $f'(u)<f(u)$. We first
claim that, after all cells of any fixed $f'$-value have been processed, the
minimum of each connected component is an unlowered vertex.

Proceed by induction over the finitely many values of $f'$. A component that
already contains a lower-level cell retains an unlowered minimum from an
earlier level. A component born at level $t$ either contains an unlowered
vertex, which precedes every lowered vertex at that level because $\prec_0$
is $f$-monotone, or initially consists only of vertices lowered to $t$. The
latter vertices form components of $L_t$ in
\Cref{lem:lowered-component-boundary}; by the end of level $t$, each is attached
either to an earlier component or to an unlowered vertex at level $t$. This
proves the claim.

We now prove \ref{prop-dd-order}. Let $x$ be a non-root vertex of $T_0'$ and
let $p'(x)$ be its parent. Since $p'(x)\prec_0'x$, either
$f'(p'(x))<f'(x)$ or the two values tie. In the first case, the component
represented by $p'(x)$ reaches a lower level, so the claim gives
\[
    f(p'(x))=f'(p'(x))<f'(x)\le f(x),
\]
and hence $p'(x)\prec_0x$. In the tie case, $p'(x)\prec_0x$ follows directly
from the compatible tie-breaking. Iteration along the parent chain proves
\ref{prop-dd-order}; its consequences for $f$ and for every
$\prec_0$-monotone reference signal are immediate.

For \ref{prop-dd-pers}, let $r$ be the $\prec_0$-minimum of $V$. Every clipping
level is an original vertex value, so $f'\ge f(r)$, while $f'\le f$ gives
$f'(r)=f(r)$. The compatible ordering therefore has the same root $r$.

Let $w\ne r$. If $w$ is lowered to $c=f'(w)$, then
the claim gives a distinct, earlier minimum in its
component by the end of level $c$. Thus its new interval dies at its birth
level and $\pers_{\tree_0'}(w)=0$.

If $w$ is unlowered, its birth value remains $f(w)$. Put
$D=f(\lk_0(w))=\fmax(m_0(w))$. Every cell in the old component through the
merge edge $m_0(w)$ has $f'$-value at most its old value and hence at most
$D$. Therefore $w$ and $p_0(w)$ are connected after level $D$ in the new
ordering. Moreover,
\[
    f'(p_0(w))\le f(p_0(w))\le f(w)=f'(w),
\]
and equality is resolved by $p_0(w)\prec_0w$, so
$p_0(w)\prec_0'w$. The new death level $d'$ of $w$ is consequently at most
$D$, and
\[
    \pers_{\tree_0'}(w)=d'-f(w)
    \le D-f(w)=\pers_{\tree_0}(w).
\]
This also covers all non-root unlowered vertices and completes the proof.
\end{proof}

\subsection*{Monotonicity under fixed tie-breaking}

The following proposition establishes the monotonicity used in the
\(\ell_\infty\) estimate for \Cref{thm:lpf-all}.

\begin{proposition}\label{prop:fixed-tie-monotonicity}
Let \(g_k=\lpf_*^{\e,k}f\), put \(h_k=\ind{(g_k)}\), and let
\(\tree_0^k=(T_0^k,\lk_0^k)\) and
\(\tree_1^k=(T_1^k,\lk_1^k)\) be the BHTs obtained with the fixed
tie-breaking scheme of \Cref{rem:lpf-choice-supp}.
Then, for every \(k\),
\begin{align}
    w\preceq_{T_0^k}u&\implies h_0(w)\le h_0(u),
        \label{eq:history-primal-monotone}\\
    w\preceq_{T_1^k}u&\implies h_0(w)\ge h_0(u).
        \label{eq:history-dual-monotone}
\end{align}
\end{proposition}

\begin{proof}
Let \(\prec_0,\prec_1\) be the initial admissible orderings for \(h_0,-h_0\).
At step \(k\), let \(\prec_0^k\) order the current values \(h_k\), breaking
ties by \(\prec_0\); define \(\prec_1^k\) analogously for \(-h_k\).

For each \(k\), let \(P_k\) denote the following assertion: if \(p\) is the
\(\prec_0^k\)-minimum of a prefix component \(C\),
then \(h_0(p)\le h_0(x)\) for every vertex \(x\in C\).
Let \(Q_k\) be the dual statement for \(\prec_1^k\), with the inequality
reversed. Both hold at \(k=0\) by value-monotonicity of the initial orderings.

Assume \(P_k,Q_k\). We prove \(P_{k+1}\); replacing \(h\) by \(-h\) gives
\(Q_{k+1}\).

For a \(0\)-LPF step, \Cref{lem:lpf-basin-decomposition} expresses the changed
vertices as disjoint prefix components \(B_i\), with BHT minima \(v_i\), each
raised to \(c_i=h_k(\lk_0^k(v_i))\). By \(P_k\), \(v_i\) minimises \(h_0\) on
\(B_i\). If a new prefix component \(C'\) lies in one \(B_i\), all its current
values tie, so its \(\prec_0^{k+1}\)-minimum is its \(\prec_0\)-minimum and
therefore minimises \(h_0\), since \(\prec_0\) is \(h_0\)-monotone. Otherwise replace every nonempty
\(C'\cap B_i\) by \(v_i\). The resulting vertices lie in one old prefix
component with the same relevant minimum: raising only delays cells of \(B_i\)
to \(c_i\), ties there retain the reference order, and every earlier outside
cell already belongs to the old prefix. Applying \(P_k\) to that component and
then using the minimality of each \(v_i\) proves \(P_{k+1}\).

For a \(1\)-LPF step, \Cref{lem:lowered-component-boundary} shows that every
connected component of vertices lowered to a common value is joined, at that
level, either to an unlowered vertex or to a component already present below
it. Hence a new primal prefix component cannot have a minimum supported only
on lowered vertices. Replace each such set of vertices by the old prefix
component reached through the joining edge.
At the old level of each removed vertex, the attachment edge is present, so
\(P_k\) compares the old component minimum with that vertex as well as with
the unchanged part. The resulting component therefore has the same minimum
for the reference values, proving \(P_{k+1}\).

Induction proves \(P_k,Q_k\) for all \(k\). If
\(w\preceq_{T_0^k}u\), then \(w\) is the minimum of a prefix component
containing \(u\), so \(P_k\) gives
\eqref{eq:history-primal-monotone}. Applying \(Q_k\) to the dual tree gives
\eqref{eq:history-dual-monotone}.
\end{proof}

\begin{lemma}\label{lem:stability}
For any \(\Gs=(\G,f)\), threshold \(\e>0\), and \(n\in\{0,1\}\),
\[
    \lpf_n^\e f\lowerpd f.
\]
Moreover,
\[
    \lpf_n^\e f\ne f
    \quad\Longleftrightarrow\quad
    |\PD_n(\G,\lpf_n^\e f)_{>0}|
    <|\PD_n(\G,f)_{>0}|.
\]
\end{lemma}

\subsection*{Proof of \Cref{lem:stability}}

We first identify the finite birth--death pairs under the induced-graph
construction. Extend an admissible ordering of \((\G,f)\) to \(G[\emb]\) as in
\Cref{rem:ordering-extension}, and use the admissible reversed ordering for
\(-\indf\).
\begin{itemize}
    \item The finite \(0\)-dimensional pairs are indexed by the BHT pairs of
    \(\indf\) born at vertices \(v\in V\). A merge edge in \(E\) represents
    the same death edge of the \gwf, while an edge \((\face,\cdot)\in E_F\)
    represents the death face \(\face\). Merges through \(E_{\hole}\) are
    essential. Pairs born at face vertices are zero-persistence formal pairs,
    and hole vertices contribute only the cases described in
    \Cref{rem:extended-real-hole}.

    \item By \eqref{eq:pd1}, the finite \(1\)-dimensional pairs are indexed by
    the BHT pairs of \(-\indf\) born at face vertices. Indeed, a complementary
    region disappears when its last cell enters, and this cell is a face:
    every boundary vertex or edge is flanked by a face of that region entering
    no earlier. The face is therefore the death cell of the \gwf\ pair, while
    the dual merge edge represents its birth.
    Pairs born at original vertices are zero-persistence: an incident face or
    hole precedes the vertex in the admissible reversed ordering and merges it
    immediately. Hole vertices again account only for essential or formal
    pairs.
\end{itemize}
Thus \(\BD(\G,\prec)\), including its zero-persistence finite-death pairs, is
indexed by primal BHT vertices in \(V\) and dual BHT vertices in \(F\); these
are precisely the representative birth cells in dimension \(0\) and
representative death cells in dimension \(1\).

For \(n\in\{0,1\}\), put
\[
    \eta_n=(-1)^n\indf,\qquad g_n=\lpf_n^\e f,\qquad
    \eta_n'=(-1)^n\ind{(g_n)}.
\]
Let \(\tree_n\) be the BHT of \(\eta_n\), and let
\(\widehat{\tree}_n\) be the BHT of \(-\eta_n\).
The definition of the two filters and
\Cref{lem:induced-lpf-consistency} give
\begin{equation}\label{eq:stab-realisation}
    \eta_n'={}^{\tree_n}\lpf_0^\e\eta_n .
\end{equation}

By \Cref{thm:BHT-stability}, \(\tree_n\) is also the BHT of
\(\eta_n'\) for the compatible ordering. Equation
\eqref{eq:bht-filtered-endpoints} leaves every interval of persistence at
least \(\e\) represented by \(\tree_n\) unchanged and collapses every
subthreshold interval. The representative vertex is unchanged; by the
preceding identification, this fixes the birth cell when \(n=0\) and the
death cell when \(n=1\).

For the other dimension, apply \Cref{prop:BHT-dual-decrease} with
\(f=-\eta_n\), \(\tree_1=\tree_n\), and
\(\tree_0=\widehat{\tree}_n\). It produces a BHT of
\[
    -{}^{\tree_n}\lpf_0^\e\eta_n=-\eta_n'
\]
whose interval indexed by each vertex has no greater persistence than before.
The same vertex indexing fixes the corresponding representative cells: face
death cells when \(n=0\), and original-vertex birth cells when \(n=1\).

The compatible output orderings for \(\eta_n'\) and \(-\eta_n'\) are reverses
of one another, so they come from a single admissible ordering of the filtered
\gwf. Combining the two correspondences therefore gives a
dimension-preserving bijection between the two birth--death-pair multisets.
It fixes the required representative cell and never increases persistence;
collapsed branches remain present as zero-persistence pairs. Hence
\(g_n\lowerpd' f\), and therefore \(g_n\lowerpd f\).

Finally, if \(g_n\ne f\), at least one positive finite interval represented
by \(\tree_n\) and having persistence below \(\e\) is collapsed. By
\Cref{thm:lpf-corollary}, all other positive intervals in dimension \(n\) are
retained; the number of essential intervals is fixed by the support.
Consequently,
\[
    |\PD_n(\G,g_n)_{>0}|<|\PD_n(\G,f)_{>0}|.
\]
The converse is immediate: a strict change in this cardinality is impossible
when \(g_n=f\). \qed

\subsection*{Proof of \Cref{thm:lpf-all}}

Let
\[
    g_k \deff \lpf_*^{\e,k}f
\]
be the sequence defined in \eqref{eq:alternating-filter}, with ties broken as
in \Cref{rem:lpf-choice-supp}. By \Cref{lem:stability}, each step satisfies
\(g_{k+1}\lowerpd g_k\). Hence the total number of features
\[
    N_k \deff |\PD_0(\G,g_k)_{>0}| + |\PD_1(\G,g_k)_{>0}|
\]
is nonincreasing. The proof of \Cref{lem:stability} constructs, for each step,
a bijection between the birth--death pairs in \(\BD\), including the
zero-persistence pairs, that does not increase persistence. Thus a
zero-persistence pair cannot become positive, while the number of essential
intervals is fixed by the support. If \(g_{k+1}\neq g_k\), then
\Cref{lem:stability}, with \(n=0\) or \(n=1\) according to the step, gives a
strict decrease in that dimension, and therefore \(N_{k+1}<N_k\). Since
\(N_k\) is a nonnegative integer, the sequence stabilises after finitely many
steps. This proves the existence of \(k_o\). Set
\(g\deff g_{k_o}=\lpf_*^\e f\).

\medskip
\noindent\textbf{The bound \(\|g-f\|_\infty<\e\).}
Put \(h_k=\ind{(g_k)}\) on the induced graph \(G[\emb]\). We prove that a strict raise of an original vertex never reaches \(h_0(u)+\e\); applying the same argument to the dual sequence \(-h_k\) proves the lower bound.

Suppose that a \(0\)-LPF step at time \(k\) strictly raises an original vertex
\(u\). By \eqref{eq:lpf0}, the new value is
\(h_{k+1}(u)=h_k(\lk_0^k(w))\) for some selected ancestor
\(w\preceq_{T_0^k}u\) in \(\tree_0^k\). A zero-persistence ancestor cannot
strictly raise \(u\), so \(0<\pers_{\tree_0^k}(w)<\e\).

We claim that any such selected vertex satisfies \(h_k(w)\le h_0(w)\).
Suppose, to the contrary, that \(h_k(w)>h_0(w)\). Then at some earlier
\(0\)-LPF step \(j<k\), the vertex \(w\) was strictly raised above its
original value. At that step \(w\) lay below a selected ancestor; by
\eqref{eq:bht-pers-monotone}, the branch indexed by \(w\) was itself
subthreshold. Hence \Cref{thm:lpf-0}, applied on the induced graph, collapsed
the \(0\)-dimensional interval represented by \(w\) to zero persistence at
time \(j+1\). At every subsequent step, \Cref{thm:BHT-stability} for a
\(0\)-LPF step or \Cref{prop:BHT-dual-decrease} for a \(1\)-LPF step fixes
the representative vertex and does not increase its persistence. Therefore
the interval represented by \(w\) still has zero \(0\)-persistence at time
\(k\), contradicting \(0<\pers_{\tree_0^k}(w)\). Thus
\(h_k(w)\le h_0(w)\).

By \Cref{prop:fixed-tie-monotonicity}, the ancestor relation
\(w \preceq_{T_0^k} u\) implies \(h_0(w) \le h_0(u)\). Combining these bounds gives
\begin{align*}
    h_{k+1}(u)
    &= h_k(\lk_0^k(w))
     = h_k(w) + \pers_{\tree_0^k}(w) \\
    &\le h_0(w) + \pers_{\tree_0^k}(w)
     < h_0(w) + \e
     \le h_0(u) + \e .
\end{align*}
Steps that lower \(u\), or leave it unchanged, preserve this upper bound once it has been established, and it is true at \(k=0\). Therefore \(h_k(u)<h_0(u)+\e\) for all \(k\) and all original vertices \(u\in V\). The same argument applied to \(-h_k\) and \(\tree_1^k\) gives \(h_k(u)>h_0(u)-\e\). Restricting from \(G[\emb]\) to \(V\) gives \(\|g_k-f\|_\infty<\e\) for every \(k\), and in particular for \(k=k_o\).

\medskip
\noindent\textbf{Remaining assertions.}
Once the alternating sequence has stabilised at \(g\), applying either
\(\lpf_0^\e\) or \(\lpf_1^\e\) changes no value. By \Cref{lem:stability},
there can be no positive subthreshold interval in either dimension, so
\[
    \PD_n(\G,g)_{<\e}\eqpd\emptyset,
    \qquad n=0,1.
\]
Finally, each step satisfies \(g_{k+1}\lowerpd g_k\) by \Cref{lem:stability}; since \(\lowerpd\) is transitive---being defined as a transitive closure in \Cref{def:lower-pd}---concatenating these steps gives
\[
    \lpf_*^\e f=g\lowerpd g_0=f .
\]
\qed

\endgroup

\fi

\vspace{-2em}
\begin{IEEEbiographynophoto}{Matias de Jong van Lier}
is currently a Ph.D. student at the Faculty of Mathematics, Kyushu University, Japan.
Coming from a background in algebraic topology, he is interested in applications of topology to data science.
\end{IEEEbiographynophoto}

\vspace{-2em}
\begin{IEEEbiographynophoto}{Sebasti\'an El\'{\i}as Graiff Zurita}
is currently an assistant professor at the Graduate School of Science, Kyoto University, Japan.
He received the M.S. degree in physics from the Balseiro Institute, Argentina, in 2015,
and the Ph.D. degree in mathematics from Kyushu University, Japan, in 2022.
\end{IEEEbiographynophoto}

\vspace{-2em}
\begin{IEEEbiographynophoto}{Shizuo Kaji}
is currently a professor at the Graduate School of Science, Kyoto University, Japan.
He received the Ph.D. degree in mathematics from Kyoto University, Japan, in 2007.
His research interests include topology and its applications.
\end{IEEEbiographynophoto}

\vfill

\end{document}